\documentclass[pre,aps,showkeys,showpacs,superscriptaddress,twocolumn,amsmath,amssymb]{revtex4} 
\usepackage{graphics}
\usepackage{epsfig}
\usepackage{amsmath}
\usepackage{setspace}
 
\usepackage{hyperref}

\begin{document}

\title[]{Spatial Extent of Branching Brownian Motion}

\author{Kabir Ramola}
\email{kabir.ramola@u-psud.fr}
\affiliation{Laboratoire de Physique Th\'{e}orique et Mod\`{e}les
Statistiques, UMR 8626, Universit\'{e} Paris-Sud 11 and CNRS,
B\^{a}timent 100, Orsay F-91405, France}

\author{Satya N. Majumdar}
\email{majumdar@lptms.u-psud.fr}
\affiliation{Laboratoire de Physique Th\'{e}orique et Mod\`{e}les
Statistiques, UMR 8626, Universit\'{e} Paris-Sud 11 and CNRS,
B\^{a}timent 100, Orsay F-91405, France}

\author{Gr\'egory Schehr}
\email{gregory.schehr@lptms.u-psud.fr}
\affiliation{Laboratoire de Physique Th\'{e}orique et Mod\`{e}les
Statistiques, UMR 8626, Universit\'{e} Paris-Sud 11 and CNRS,
B\^{a}timent 100, Orsay F-91405, France}

\date{\today}

\begin{abstract}
We study the one dimensional branching Brownian motion starting at the origin and 
investigate the correlation between the rightmost ($X_{\max}\geq 0$) and leftmost ($X_{\min} \leq 0$)
visited sites up to time $t$. At each time step the existing particles in the system 
either diffuse (with diffusion constant $D$), die (with rate $a$) or split into two particles (with rate $b$).
We focus on the regime $b \leq a$ where these two extreme values $X_{\max}$ and $X_{\min}$ are strongly correlated.
We show that at large time $t$, the joint probability distribution function (PDF) of the two extreme points
becomes stationary $P(X,Y,t \to \infty) \to p(X,Y)$. Our exact results for $p(X,Y)$ demonstrate that the correlation between $X_{\max}$ and $X_{\min}$ is nonzero, 
even in the stationary state. From this joint PDF, we compute exactly the stationary PDF $p(\zeta)$ of 
the (dimensionless) span $\zeta = {(X_{\max} - X_{\min})}/{\sqrt{D/b}}$, which is the distance between the rightmost and leftmost visited sites. This span distribution is characterized by a linear behavior 
${p}(\zeta) \sim \frac{1}{2} \left(1 + \Delta \right) \zeta$
for small spans, with $\Delta = \left(\frac{a}{b} -1\right)$. In the critical case ($\Delta = 0$) this distribution has a non-trivial power law tail
${p}(\zeta) \sim 8 \pi \sqrt{3} /\zeta^3$ for large spans. 
On the other hand, in the subcritical case ($\Delta > 0$), we show that the span distribution decays exponentially as
${p}(\zeta) \sim (A^2/2) \zeta \exp \left(- \sqrt{\Delta}~\zeta\right)$ for large spans, where $A$ is a non-trivial function
of $\Delta$ which we compute exactly. We show that these asymptotic behaviors carry the signatures of the correlation between
$X_{\max}$ and $X_{\min}$. Finally we verify our results via direct Monte Carlo simulations.
\end{abstract}


\maketitle

\section{Introduction}
\begin{figure}[h]
\includegraphics[width=1\linewidth]{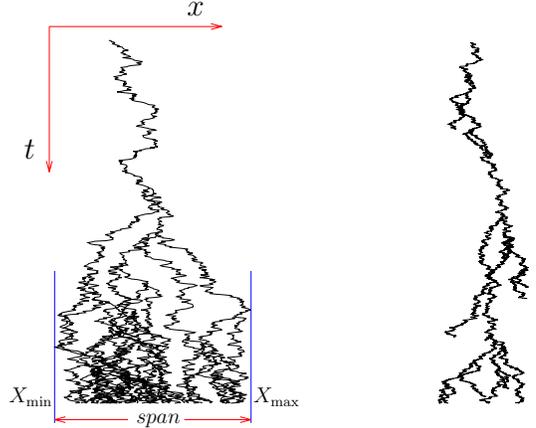}
\caption{A realization of the dynamics of branching Brownian motion (left) in the supercritical case (right) 
in the critical case. The span of the process is defined as $s = X_{\max} - X_{\min}$,  where $X_{\max}$ and $X_{\min}$
are the maximum and minimum displacements of the process up to a certain time $t$ respectively.}
\label{walk_picture}
\end{figure}

Branching Brownian motion (BBM) is a well-known model that finds applications in several
areas of science including physics, mathematics and biology.
BBM arises naturally in the context of systems where new particles are generated at each 
time step such as models of evolution, epidemiology, population growth and nuclear reactions, and now has a long history
\cite{fisher,harris,golding,sawyer,bailey,mckean,bramson,brunet_derrida_epl,brunet_derrida_jstatphys,mezard,derrida_spohn,demassi,majumdar_pnas,derrida_brunet_simon,zhuang2,zoia}.
In addition, BBM has also been widely used in theoretical physics where it has been studied in the context of
reaction-diffusion models, disordered systems amongst others \cite{demassi,derrida_spohn}. 
BBM is also an important model in probability theory as it combines the long-studied 
diffusive motion with the random branching mechanism of Galton-Watson trees \cite{galton_watson}. 
In this paper we are interested in one dimensional BBM.
The process begins with a single particle at the position $x = 0$ at time $t = 0$. 
The dynamics proceeds in continuous time, where in a small time interval $\Delta t$, each particle splits into two 
independent particles with probability $b\, \Delta t$, 
dies with with probability $a\, \Delta t$, and with the remaining probability
$(1-(a+b)\Delta t)$ performs a Brownian motion on a line with a diffusion constant $D$. 
A realization of the dynamics of such a process is shown in Fig. \ref{walk_picture}.

In a given realization of this BBM process, there are in general $N(t) \ge 0$ particles present 
in the system at a particular time $t$. The parameters $b$ and $a$ in this BBM model define three regimes with different properties.
The number of particles $N(t)$ is a random variable whose statistics depends on $a$ and $b$. 
When the rate of birth is greater than the death rate ($a < b$), the {\it supercritical} phase, 
the process is explosive and the average
number of particles in the system grows exponentially with time $\langle N(t) \rangle = \exp((b-a)t)$.
In contrast, when the birth rate is smaller than the death rate ($b < a$), the {\it subcritical} phase, the process 
eventually dies and, on  
an average, there are no particles present in the system as $t\to \infty$. At the critical point $a = b$, the system
is characterized by a fluctuating number of particles with $\langle N(t) \rangle = 1$ at all times $t$.

If one takes a snapshot of the system at a given time $t$, the spatial positions of the existing particles happen to be  
strongly correlated, since the particles are linked by their common genealogy.
One important object that has been extensively studied is the order statistics of these particles, i.e.,
the statistics of the position $x_k(t)$ of say the $k$-th rightmost particle at time $t$, where the
particle positions on the line are
ordered as $x_1(t) > x_2(t) > .... > x_{N(t)}(t)$
\cite{sawyer,mckean,bramson,brunet_derrida_epl,brunet_derrida_jstatphys,ramola_majumdar_schehr,ramola_majumdar_schehr2}. 
Another related interesting quantity is the gap $g_k(t) = x_k(t) - x_{k+1}(t)$ between the $k$-th and $(k+1)$-th particle at time $t$.   
Most of these studies have thus focused on extreme value questions at a given time $t$. However, there are other interesting extreme value observables 
that concern the history of the process over the entire time interval $[0,t]$. 
For instance, one can consider the global maximum,   
$X_{\max} = \textmd{max}_{0 \le \tau < t}\left[\{x_{1}(\tau),x_{2}(\tau),x_{3}(\tau) ...,x_{N(\tau)}(\tau) \}\right]$
which represents the maximum of all the particle positions {\it up to} time $t$. This has the simple interpretation as the maximum displacement of 
the entire process up to time $t$ (see Fig. \ref{walk_picture}). 
This global maximum has appeared in a variety of applications including the spread of gene populations~\cite{sawyer} and 
the propagation of animal epidemics in two dimensions \cite{majumdar_pnas}. Similarly the global minimum
$X_{\min} = \textmd{min}_{0 \le \tau < t}\left[\{x_{1}(\tau),x_{2}(\tau),x_{3}(\tau) ...,x_{N(\tau)}(\tau) \}\right]$  
is another
interesting quantity that, by symmetry, has the same marginal
probability distribution function (PDF) as $-X_{\max}$. 

The marginal PDF of $X_{\max}$ has been studied extensively for the supercritical \cite{mckean,bramson}, critical and the subcritical phases \cite{sawyer,iscoe}. 
While the marginal distributions of $X_{\max}$, and hence that of $-X_{\min}$, are well studied, much less is known about the correlation between these two random variables.
In this paper, we study the joint PDF of $X_{\max}$ and $X_{\min}$. 
In the supercritical phase, this joint PDF is always time dependent, and is hard to compute analytically. However, in this case, 
$X_{\max}$ and $X_{\min}$ get separated from each other ballistically in time and hence become uncorrelated at late times. 
In contrast, 
in the critical and subcritical phases ($b \leq a$), we show that the joint PDF reaches a limiting stationary form at late times, which we compute analytically.
Moreover, for $b\leq a$, our exact results for the stationary joint PDF demonstrate that this correlation between $X_{\max}$ and $X_{\min}$ remains finite even in the stationary state. 

The joint PDF of $X_{\max}$ and $X_{\min}$ has the following interesting physical application. For instance, 
in the context of epidemic spreads, it is important to characterize the spatial extent over which the epidemic has propagated
up to time $t$. This is clearly measured by the span $s = X_{\max} - X_{\min}$ of the process up to time $t$ (see Fig. \ref{walk_picture}) \cite{larralde,vishwanathan,kundu}.
Evidently, to compute the distribution of $s$, we need to know the joint PDF of $X_{\max}$ and $X_{\min}$. 
In this paper we also compute analytically the stationary PDF of the span $s$ in the critical ($b=a$) and the subcritical ($b<a$) cases. 
Our exact results demonstrate that the correlation between  $X_{\max}$ and $X_{\min}$ is also manifest in the stationary span PDF.

The rest of the paper is organized as follows. In section II, we define the model precisely and summarize our main results.  
In Section III we derive an exact evolution equation for the
joint distribution of $X_{\max}$ and $X_{\min}$. 
In Section IV we derive the stationary
joint PDF of $X_{\max}$ and $X_{\min}$ for the critical ($b=a$) and the subcritical ($b<a$) cases. In Section V we compute
the stationary PDF of the span and extract its asymptotic behaviors analytically. 
In Section VI we compare our analytical predictions with Monte Carlo simulations. Finally, we conclude with a discussion
in Section VII. Some details of computations are relegated to the appendices. 

\section{The model and a summary of the results}\label{section:results}

\begin{figure}
\centering
\includegraphics[width = 1\linewidth]{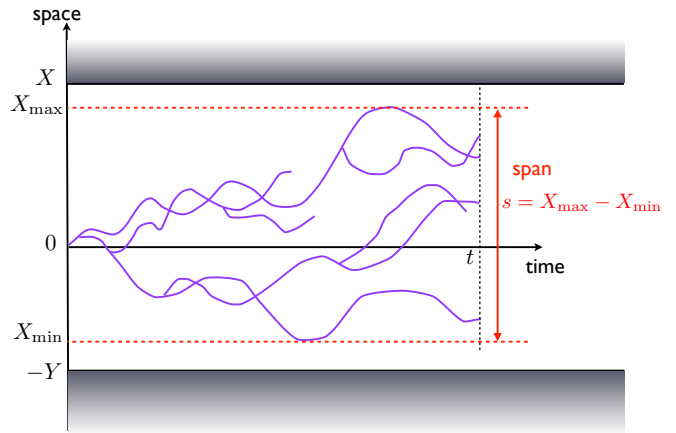}
\caption{Schematic representation of a trajectory of the BBM confined in the box $[-Y,X]$. Note that $X_{\max}$ and $X_{\min}$ denote respectively 
the maximum and the minimum of the process {\it up to time} $t$. 
The process starts with a single particle at the origin at time $t=0$ and hence $X_{\max} \geq 0$ while $X_{\min} \leq 0$.}\label{fig:cartoon}
\end{figure}

\noindent{\bf The model and the observables.} We consider the BBM on a line starting with a single particle at the 
origin at time $t=0$. The process evolves via the following continuous time dynamics. In a small time interval 
$\Delta t$, each existing particle (i) dies with probability $a \, \Delta t$, (ii) branches into two offspring with 
probability $b \, \Delta t$ and (iii) diffuses, with diffusion constant $D$, with the remaining probability $1 - 
(a+b)\, \Delta t$. A schematic trajectory of the process is shown in Fig. \ref{fig:cartoon}, where $X_{\max}$ and 
$X_{\min}$ denote respectively the maximal displacements of the process up to time $t$ in the positive and the 
negative direction. It is convenient to define these observables in their dimensionless forms $x_{\max} 
= X_{\max}/\sqrt{D/b}$ and $x_{\min} = X_{\min}/\sqrt{D/b}$. Since the particle starts at the origin, $x_{\max} \geq 
0$ necessarily and similarly $x_{\min} \leq 0$ necessarily. In the subsequent discussions we find it convenient to consider the positive quantities
$x_{\max}$ and $-x_{\min}$ as our basic random variables.

As mentioned in the introduction, the marginal PDF of $x_{\max}$ (and consequently that of $-x_{\min}$),
 has been extensively studied for all $a$ and $b$ \cite{sawyer,mckean,bramson}.
While in the supercritical phase ($b>a$), this marginal PDF remains time dependent for all $t$ \cite{mckean,bramson}, 
for $b \leq a$, it approaches a stationary form $p_{\rm marg}(x)$ which is known explicitly. 
It is convenient to express it in terms of its cumulative distribution ${\cal R}(x) = \int_{x}^\infty p_{\rm marg}(x') dx'$. 
We set $\Delta = a/b-1$. In the critical case $\Delta = 0$~\cite{sawyer}, 
\begin{equation}
\mathcal{R}(x) = \frac{1}{\left(1 + \dfrac{x}{\sqrt{6}} \right)^2} \;. 
\label{max_distribution_critical}
\end{equation}
Consequently, $p_{\rm marg}(x) = -d\mathcal{R}(x)/dx$ has the asymptotic behaviors
\begin{eqnarray}\label{pmarg_asympt_critical}
p_{\rm marg}(x) \sim
\begin{cases}
p_{\rm marg}(0) = \sqrt{\dfrac{2}{3}} \;, \; x \to 0 \\
\\
\dfrac{12}{x^3} \;, \; x \to \infty \;.
\end{cases}
\end{eqnarray}
In the subcritical case $\Delta > 0$~\cite{sawyer}: 
\begin{equation}\label{max_distribution_subcritical_0}
\mathcal{R}(x) =  \frac{3 \Delta}{2} \textmd{csch}^2\left(\frac{\sqrt{\Delta}}{2} x  + 
\sinh^{-1}\sqrt{\frac{3 \Delta}{2}}\right).
\end{equation}
This result can further be simplified to give
\begin{eqnarray}\label{max_distribution_subcritical}
\mathcal{R}(x) = \frac{\alpha - 1}{[-1 + \alpha \cosh(\sqrt{\Delta} x) + \sqrt{(\alpha^2-1)} \sinh{(\sqrt{\Delta} x)} ]} 
\end{eqnarray}
where $\alpha = 1 + 3 \Delta$. 

Consequently, $p_{\rm marg}(x) = -d\mathcal{R}(x)/dx$ has the asymptotic behaviors
\begin{eqnarray}
\label{pmarg_asympt_subcritical}
p_{\rm marg}(x) \sim
&&\hspace{-0.45cm}
\begin{cases}
p_{\rm marg}(0)= \sqrt{\dfrac{2}{3} + \Delta} \;, \; x \to 0 \\ \\
6 \Delta^{\frac{3}{2}} e^{-2 \sinh^{-1} \sqrt{\frac{3\Delta}{2}}} \exp{(-\sqrt{\Delta}\, x)} \;, \; 
x \to \infty \;.  
\end{cases}
\end{eqnarray}

By symmetry, $-x_{\min}$ has the same marginal PDF $p_{\rm marg}(x)$ for $\Delta \geq 0$. While $p_{\rm marg}(x)$ is thus well known, 
in this paper we compute the joint stationary PDF $p(x,y)$ of $x = x_{\max}$ and $y = -x_{\min}$ for $\Delta \geq 0$. 
One of our main results is to highlight the nonzero correlation between $x_{\max}$ and $-x_{\min}$ even in the stationary state. Indeed we show that 
\begin{equation}
\label{hypo.1}
p(x,y) \neq p_{\rm uncorr}(x,y) = p_{\rm marg}(x) p_{\rm marg}(y) \;.
\end{equation}
The stationary PDF of the dimensionless span $\zeta = s/\sqrt{D/b} = x_{\max} - x_{\min}$ can be computed from the joint PDF $p(x,y)$ via the relation
\begin{equation}
\label{hypo.2}
p(\zeta) = \int_0^\infty dx \int_0^\infty dy \, p(x,y) \, \delta(x+y-\zeta) \;.
\end{equation} 
We compute $p(\zeta)$ exactly for all $\Delta \geq 0$ and find the following asymptotic behaviors. 


\noindent {\bf Critical case ($\Delta = 0$):} In this case we find
\begin{eqnarray}\label{pzeta_asympt_critical}
p(\zeta) \sim
\begin{cases}
\dfrac{\zeta}{2} \;, \;\; \zeta \to 0 \\
\\
\dfrac{{\cal A}}{\zeta^3} \;, \;\; \zeta \to \infty \;, \: {\cal A} = 8 \pi \sqrt{3} = 43.53118\ldots
\end{cases}
\end{eqnarray}

\vspace*{0.3cm}

\noindent {\bf Subcritical case ($\Delta > 0$):} Here we get
\begin{eqnarray}\label{pzeta_asympt_subcritical}
p(\zeta) \sim
\begin{cases}
\dfrac{1}{2}\left( 1 + \Delta \right)\zeta \;, \;\; \zeta \to 0 \\
\\
\dfrac{A^2}{2} \, \zeta \, \exp{(-\sqrt{\Delta}\,\zeta)}  \;, \;\; \zeta \to \infty \;.
\end{cases}
\end{eqnarray}
where $A= 12 \, \Delta \, \left[\sqrt{3\Delta/2} + \sqrt{1+3\Delta/2}\right]^{-2}$.

\noindent{\bf Signatures of the correlation between $x_{\max}$ and $x_{\min}$.} Interestingly, 
one can show that these asymptotic behaviors of $p(\zeta)$ for the critical (\ref{pzeta_asympt_critical}) and the subcritical cases (\ref{pzeta_asympt_subcritical})
 carry the signatures of the correlation between $x_{\max}$ and $x_{\min}$ (see also Figs. \ref{fig:stationary_dist} and \ref{fig:stationary_dist_subcritical} below).
 In order to demonstrate this, we compute the asymptotic behaviors of $p(\zeta)$ in the
hypothetical case where one assumes that $x_{\max}$ and $x_{\min}$ are completely uncorrelated. 
Given that $x_{\max}$ and $-x_{\min}$ have the same PDF $p_{\rm marg}(x)$ [obtained from
 Eq. (\ref{max_distribution_critical}) for $\Delta = 0$ and from Eq. (\ref{max_distribution_subcritical}) for $\Delta > 0$], 
the span PDF $p_{\rm uncorr}(\zeta)$, assuming that $x_{\max}$ and $x_{\min}$  are uncorrelated can be obtained by inserting 
$p(x,y) = p_{\rm uncorr}(x,y)$ into Eq. (\ref{hypo.2}) and 
is given~by 
\begin{equation}\label{uncorr_marginal}
p_{\rm uncorr}(\zeta) = \int_0^\zeta p_{\rm marg}(x) p_{\rm marg}(\zeta - x) dx  \;.
\end{equation} 
For small $\zeta$, it behaves as 
\begin{equation}\label{uncorr_small}
p_{\rm uncorr}(\zeta) \sim  p^2_{\rm marg}(0) \,\zeta \;, {\rm when}\; \zeta \to 0 \;.
\end{equation}
Substituting $p_{\rm marg}(0) =  \sqrt{2/3+\Delta}$ from Eq. (\ref{pmarg_asympt_subcritical}) 
in Eq. (\ref{uncorr_small}) gives 
\begin{equation}\label{uncorr_small_final}
p_{\rm uncorr}(\zeta) \sim  \left(\frac{2}{3} + \Delta\right) \,\zeta \;, {\rm when}\; \zeta \to 0 \;.
\end{equation}
Comparing this result with the exact one in Eqs. (\ref{pzeta_asympt_critical}) and (\ref{pzeta_asympt_subcritical}), we see that,
 while both of them grow linearly for small $\zeta$, the slopes are different, reflecting the fact that $x_{\rm max}$ and $x_{\rm min}$ are actually correlated.

To investigate the large $\zeta$ behavior of $p_{\rm uncorr}(\zeta)$ in Eq. (\ref{uncorr_marginal}), 
we need to treat separately the critical ($\Delta = 0$) and the subcritical ($\Delta > 0$) cases -- see 
Eqs. (\ref{pmarg_asympt_critical}) and (\ref{pmarg_asympt_subcritical}). 
In the critical case ($\Delta = 0$), substituting the asymptotic behavior 
$p_{\rm marg}(x) \sim 12/x^3$ from Eq. (\ref{pmarg_asympt_critical}) in Eq. (\ref{uncorr_marginal}), one gets for large $\zeta$
\begin{equation}
p_{\rm uncorr}(\zeta) \sim \frac{24}{\zeta^3} \;, \; {\rm for} \; \Delta = 0 \;.
\label{uncorr_large_critical}
\end{equation}
While this uncorrelated assumption correctly reproduces the $\zeta^{-3}$ decay 
(\ref{pzeta_asympt_critical}), the prefactor $24$ is different from the exact value 
${\cal A} = 8\pi\sqrt{3} =  43.53118\ldots$ in Eq. (\ref{pzeta_asympt_critical}), 
again reflecting the nonzero correlation between $x_{\max}$ and $x_{\min}$. On the other hand, for $\Delta > 0$, 
one obtains from Eqs. (\ref{max_distribution_subcritical}) and (\ref{uncorr_marginal}):
\begin{equation}\label{uncorr_large_final}
p_{\rm uncorr}(\zeta) \sim \frac{\Delta}{4} A^2 \, \zeta \exp{\left(-\sqrt{\Delta} ~\zeta\right)} \;, {\rm for} \; \Delta > 0 \;.
\end{equation} 
Here also, the assumption of vanishing correlation correctly reproduces the $\zeta$-dependence 
$\propto \zeta  \exp{\left(-\sqrt{\Delta} ~\zeta\right)}$ of the right tail (\ref{pzeta_asympt_subcritical}) 
but the amplitude is incorrect by a factor $\Delta/2$, reflecting once again the presence of
finite correlations between $x_{\max}$ and $x_{\min}$.

\section{Joint distribution of the maximum and minimum}

We are interested in the spatial extent of the BBM process {\it up to} time $t$. 
The process begins with a single particle at $x =0$ at time $t =0$. 
We recall that the span of the process up to $t$, characterizing the spatial extent, is defined as $s = X_{\max} - X_{\min}$,  
where $X_{\max}$ and $X_{\min}$
are respectively the maximum and minimum displacements of the process up to time $t$ 
(see Fig. \ref{walk_picture}).

We start by defining the joint cumulative probability 
\begin{equation}
Q(X,Y,t) \equiv \Pr \textmd{\,\{$X_{\max} < X$, $X_{\min} > -Y$; up to time $t$\}}\;. \nonumber
\end{equation}
This has the simple interpretation as the probability that the process is confined
within the box $[-Y,X]$ {\it up to} time $t$ (see Fig. \ref{fig:cartoon}).
The marginal cumulative distribution of the maximum can be obtained by taking the
$Y \to \infty$ limit of $Q(X,Y,t)$. Similarly the marginal cumulative distribution of the minimum is obtained by taking the
$X \to \infty$ limit. The joint PDF $P(X,Y,t)$ of $X_{\max}=X$ and $-X_{\min}=Y$ is then given by
\begin{equation}
P(X,Y,t) = \frac{\partial }{\partial X} \frac{\partial}{\partial Y} Q(X,Y,t)
\label{joint_pdf_unscaled}
\end{equation}
where, by definition, $X\ge 0$ and $Y\ge 0$.
The PDF of the span $s=X+Y$ is then given by
\begin{equation}
P(s,t) = \int_{0}^{\infty} \int_{0}^{\infty} dX dY \delta(X + Y - s) P(X,Y,t). 
\label{span_definition}
\end{equation}
\subsection{Backward Fokker-Planck equation for $Q(X,Y,t)$}
\label{bfp_Q_derive}

We derive a backward Fokker-Planck (BFP) equation for $Q(X,Y,t)$, following similar steps as in Refs.~\cite{ramola_majumdar_schehr,ramola_majumdar_schehr2}. 
We investigate how $Q(X,Y,t)$ evolves into $Q(X,Y,t+\Delta t)$. The
goal is to derive a differential equation for the evolution of $Q(X,Y,t)$. For this purpose we split the time interval $[0,t +\Delta t]$ into two
subintervals: $[0,\Delta t]$ and $[\Delta t, t+\Delta t]$. We then take into account all possible stochastic events that take place in the first subinterval $[0,\Delta t]$. 
In $[0,\Delta t]$, the particle at $x=0$ can\\

{\bf A)} split into two particles with probability $b \Delta t$, resulting in two BBM processes that are both confined 
within $[-Y,X]$ up to time $t + \Delta t$ with probability $Q^2(X,Y,t)$. The contribution from this term to $Q(X,Y,t + \Delta t)$ is then
$b \Delta t ~Q^2(X,Y,t)$.\\ 

{\bf B)} die with a probability $a \Delta t$, leading to no particles at subsequent times. This event automatically ensures 
with probability $1$ that the process remains confined within $[-Y,X]$ up to $t +\Delta t$. Hence it contributes a term $a \Delta t \times 1$ to $Q(X,Y,t + \Delta t)$. \\

{\bf C)} diffuse with probability $1-(b+a)\Delta t$, moving a distance $\Delta x = \eta(0)\Delta t$ in the first time step. 
This shifts the process by a distance $\Delta x$ at the first time step. The probability that the resulting process is 
confined within $[-Y,X]$ up to time $t + \Delta t$
is then given by $\langle Q(X - \Delta x, Y + \Delta x, t) \rangle_{\eta(0)}$. 
By the subscript ${\eta(0)}$ we denote an averaging over all possible values of the diffusive jump at the first time step.
Hence this term contributes $\left(1- (b+a)\Delta t\right) \times \langle Q (X - \Delta x, Y+ \Delta x,t) \rangle_{\eta(0)}$ to the final probability
$Q(X,Y,t+\Delta t)$.\\

Adding the contributions from these three terms {\bf A)}, {\bf B)} and {\bf C)}, we have
\begin{eqnarray}
\nonumber
Q(X,Y,t + \Delta t) &=&  b \Delta t ~Q^2(X,Y,t)+ a \Delta t\\
&&\hspace{-0.32\linewidth}+\left(1- (b+a)\Delta t\right) \langle Q (X - \Delta x, Y+ \Delta x,t) \rangle_{\eta(0)},
\label{focker_planck_1}
\end{eqnarray}
where $\eta(t)$ is a Gaussian white noise process with the properties
\begin{flalign}
\langle \eta(t) \rangle =0 \;, \;  \langle \eta(t)\eta(t') \rangle = 2 D \delta(t - t').
\label{noise_def}
\end{flalign}
Taylor expanding Eq. (\ref{focker_planck_1}), using the properties of the noise in Eq. (\ref{noise_def}), 
and taking the limit $\Delta t \to 0$, we arrive at the exact BFP evolution equation
\begin{eqnarray}
\nonumber
\frac{\partial}{\partial t} Q(X,Y,t) = D\left( \frac{\partial}{\partial X} -  
\frac{\partial}{\partial Y}\right)^2  Q(X,Y,t) + a\\
-(b+a) ~Q(X,Y,t) + b ~Q^2(X,Y,t).
\label{bfp_Q}
\end{eqnarray}
Since at time $t=0$, both the maximum and minimum of the process is at $x=0$, the initial condition
is
\begin{equation}
Q(X,Y,0) = \Theta(X)\Theta(Y) \;,
\end{equation}
where $\Theta$ is the Heaviside step function defined as
\begin{equation}
\Theta(x) = 
\begin{cases}
1 ~~~~~~~\textmd{for}~~~~~~~  x>0, \\
0 ~~~~~~~\textmd{for}~~~~~~~  x<0 \;.
\end{cases}
\label{heaviside} 
\end{equation}
At any time $t > 0$, the maximum $X_{\max} \ge 0$ and 
the minimum $-X_{\min} \ge 0$, leading to the boundary conditions 
\begin{equation}
Q(X,Y,t) =
\begin{cases}
0 ~~~~~~~\textmd{for}~~~~~~~  X<0, \\
0 ~~~~~~~\textmd{for}~~~~~~~  Y<0 \;.
\end{cases}
\label{Q_boundaries}
\end{equation}
It is actually convenient to work with  
\begin{equation}
R(X,Y,t) = 1- Q(X,Y,t) \;,
\label{R_def}
\end{equation}
which denotes the complementary probability that the maximum or minimum up to time $t$
is {\it not} within $[-Y,X]$.
Inserting Eq. (\ref{R_def}) into Eq. (\ref{bfp_Q}) we have
\begin{eqnarray}
\nonumber
\frac{\partial R(X,Y,t)}{\partial t} = D\left( \frac{\partial}{\partial X} -  
\frac{\partial}{\partial Y}\right)^2 R(X,Y,t)\\ 
\hspace*{0.5cm}+ (b - a)~R(X,Y,t) - b ~R^2(X,Y,t), 
\label{FKPP_R}
\end{eqnarray}
with the initial conditions
\begin{equation}
R(X,Y,0) = \Theta(-X)\Theta(-Y) \;,
\end{equation}
and the boundary conditions
\begin{equation}
R(X,Y,t) =
\begin{cases}
1 ~~~~~~~\textmd{for}~~~~~~~ X<0, \\
1 ~~~~~~~\textmd{for}~~~~~~~ Y<0.
\end{cases}
\label{R_boundaries}
\end{equation}

\subsection{Dimensionless Variables}
It is natural to consider the evolution equations in terms of dimensionless variables as follows
\begin{eqnarray}
\nonumber
x = \frac{X}{\sqrt{D/b}},\\
\nonumber
y = \frac{Y}{\sqrt{D/b}},\\
\nonumber
\tau = b t,\\
\Delta = \frac{a}{b} - 1.
\label{rescaled_def}
\end{eqnarray}
Similarly, we can define the dimensionless span of the process as
\begin{equation}
\zeta = x + y = \frac{s}{\sqrt{D/b}} = \frac{X_{\max} - X_{\min}}{\sqrt{D/b}}.
\label{dimensionless_span_def} 
\end{equation}
Our goal in this paper is to derive the stationary joint PDF of $x$ and $y$ and also
the stationary PDF of $\zeta$. 
In order to avoid a proliferation of symbols, we keep the same
notation for the PDF's of the
unscaled and scaled variables, with $P(X,Y,t) \to P(x,y,\tau)$ and $P(s,t) \to P(\zeta,\tau)$.
Similarly we have $R(X,Y,t) \to R(x,y,\tau)$. 
The distributions of the scaled variables are related to the unscaled distributions as
\begin{eqnarray}
\nonumber
&&P(\zeta,\tau) = \frac{1}{\sqrt{D/b}} P\left(\zeta = \frac{s}{\sqrt{D/b}},\tau=bt\right),\\
\nonumber
&&P(x,y,\tau) = \frac{1}{D/b} P\left(x=\frac{X}{\sqrt{D/b}},y=\frac{Y}{\sqrt{D/b}},\tau=bt\right).\\
\label{rescaled_dist}
\end{eqnarray}
In terms of these scaled variables Eq. (\ref{FKPP_R}) takes the simpler form
\begin{eqnarray}
\nonumber
\frac{\partial R(x,y,\tau)}{\partial \tau} &=& \left( \frac{\partial}{\partial x} -  
\frac{\partial}{\partial y}\right)^2 R(x,y,\tau)\\
&&- \Delta R(x,y,\tau) -  R^2(x,y,\tau).
\label{FKPP_R_rescaled} 
\end{eqnarray}
Eq. (\ref{FKPP_R_rescaled}) is a non-linear equation whose explicit solution at finite time $t$ 
is hard to obtain analytically. In the supercritical case $b>a$, we expect this solution
to be time dependent at all times $t$. However, for $b\le a$, we show below that as $\tau\to \infty$,
Eq. (\ref{FKPP_R_rescaled}) admits a stationary solution for $R(x,y,\tau)$ that
can be computed explicitly.
Using this solution, and Eqs. (\ref{joint_pdf_unscaled}), (\ref{R_def}) and (\ref{rescaled_def}), 
the joint PDF of $x$ and $y$ can then be expressed as
\begin{equation}
P(x,y,\tau) = - \frac{\partial}{\partial x} \frac{\partial}{\partial y} R(x,y,\tau).
\label{xy_jointPDF}
\end{equation}
Finally, this joint PDF can be used to evaluate the PDF of the dimensionless span of this process
 defined in Eq. (\ref{span_definition}), which is then given by
\begin{equation}
{P}(\zeta,\tau) = \int_{0}^{\infty} \int_{0}^{\infty} dx dy \delta(x + y - \zeta) {P}(x,y,\tau). 
\label{span_definition2}
\end{equation}
At large times, this PDF converges to the stationary distribution 
${p}(\zeta) = {P}(\zeta,\tau \to \infty)$, which we analyse in detail in section \ref{section:span}.

\section{Stationary Joint Distribution of $x_{\max}$ and $x_{\min}$ for $b \leq a$}

\begin{figure}[ht]
\begin{center}
\vspace*{0.cm}
\includegraphics[width=1\linewidth]{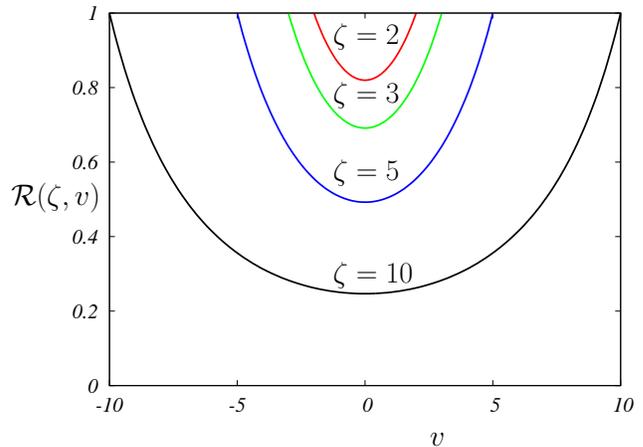}
\end{center}
\caption{${\cal R}(\zeta,v)$ as a function of $v \in [-\zeta,+\zeta]$ for different values of $\zeta$. For fixed $\zeta$,
${\cal R}(\zeta,v)$ is a smooth non-monotonic function, 
symmetric around 
$v=0$ in $-\zeta \leq v \leq + \zeta$, and has a minimum at $v=0$.
The data plotted above corresponds to the case $\Delta = 0$ and was 
obtained by numerically evaluating Eqs. (\ref{Rstar_equation_critical}) and (\ref{R_equation}).}
\label{Rzetav_fig}
\end{figure}

We now focus on the case $b\leq a$ where the joint distribution $R(x,y,\tau)$ in Eq. (\ref{FKPP_R_rescaled}) is
expected to approach a stationary limit as $\tau \to \infty$: 
\begin{equation}
\mathcal{R}(x, y) = {R}(x, y, \tau \to \infty).
\end{equation}
Setting the left hand side (lhs) of Eq. (\ref{FKPP_R_rescaled}) to $0$ in the stationary limit gives
\begin{equation}
\left( \frac{\partial}{\partial x} -  \frac{\partial}{\partial y}\right)^2 
\mathcal{R}(x,y)  = \Delta \mathcal{R}(x,y) + \mathcal{R}^2(x,y) \;, 
\label{FKPP_subcritical2}
\end{equation}
for $\; x \geq 0 \;, y \geq 0$.
\begin{figure}[ht]
\begin{center}
\vspace*{0.cm}
\includegraphics[width=0.6\linewidth]{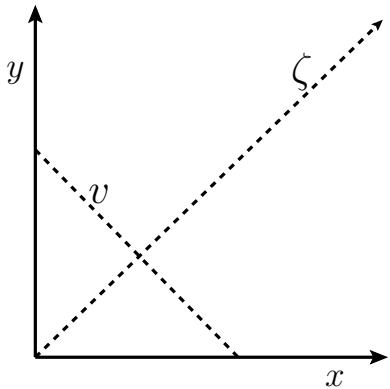}
\end{center}
\caption{The change of variables $\{x,y\} \rightarrow \{\zeta,v\}$.}
\label{variable_change_fig}
\end{figure}
Next, it is convenient to make a change of variables
\begin{eqnarray}
\nonumber
\zeta = x + y,\\
v = x - y,
\label{change_of_variables}
\end{eqnarray}
with $\zeta \in [0,\infty)$ and $v \in [-\zeta,\zeta]$ (see Fig. \ref{variable_change_fig}). Note that the variable $\zeta$ represents
the dimensionless span of the process.
In terms of these new variables Eq. (\ref{FKPP_subcritical2}) becomes
\begin{equation}
4 \left( \frac{\partial}{\partial v}\right)^2 
\mathcal{R}(\zeta,v)  - \Delta \mathcal{R}(\zeta,v) - \mathcal{R}^2(\zeta,v) = 0 \;,
\label{FKPP_subcritical3}
\end{equation}
valid in the regime $v \in [-\zeta, +\zeta]$ and $\zeta \in [0,+\infty)$ (see Fig. \ref{variable_change_fig}).
When $v \to + \zeta$, i.e., $y \to 0$, the boundary condition $R(x,y=0,t) = 1$ given in Eq. (\ref{R_boundaries}), 
translates into ${\mathcal R}(\zeta,\zeta) = 1$. Similarly, when $v \to - \zeta$,  i.e., $x \to 0$, 
the boundary condition $R(x=0,y,t) = 1$ translates into ${\mathcal R}(\zeta,-\zeta)  = 1$. 
In addition, the solution must be symmetric around $v=0$ (corresponding to $x=y$). Since ${\mathcal R}(\zeta,v)$ 
is a cumulative probability, $0 \leq {\mathcal R}(\zeta,v) \leq 1$.  
Consequently, for a fixed $\zeta$, as $v$ decreases from $\zeta$ we expect that ${\mathcal R}(\zeta,v)$ should decrease 
from its value ${\mathcal R}(\zeta,\zeta) = 1$. 
By symmetry, as $v$ increases from $-\zeta$, ${\mathcal R}(\zeta,v)$ should decrease from its value ${\mathcal R}(\zeta,-\zeta) = 1$. 
Thus we expect ${\mathcal R}(\zeta,v)$, as a function of $v$ for fixed $\zeta$, is a smooth non-monotonic function, 
symmetric around 
$v=0$ in $-\zeta \leq v \leq + \zeta$, and with a minimum at $v=0$ (see Fig. \ref{Rzetav_fig}). 
Assuming analyticity around the minimum at $v=0$ gives the condition
\begin{equation}\label{condition}
\frac{\partial \mathcal{R}(\zeta,v)}{\partial v}\Big\lvert_{v = 0} = 0.
\end{equation}  

Once we find the solution $\mathcal{R}(\zeta,v) $ of Eq. (\ref{FKPP_subcritical3}), using Eqs. (\ref{xy_jointPDF}) and (\ref{change_of_variables}) 
the joint PDF of $\zeta$ and $v$ can be expressed as 
\begin{equation}
{p}(\zeta,v)  =  -\frac{1}{2}\frac{\partial}{\partial x} \frac{\partial}{\partial y} \mathcal{R}(x,y) \equiv  \frac{1}{2}
\left(\frac{\partial^2}{\partial v^2} -  \frac{\partial^2}{\partial \zeta^2} \right) \mathcal{R}(\zeta,v).
\label{joint_PDF_sv}
\end{equation}
where the factor $1/2$ in (\ref{joint_PDF_sv}) comes from the Jacobian of the transformation $\{x,y\} \to \{\zeta,v\}$ (\ref{change_of_variables}) such that 
\begin{eqnarray}\label{normalization_joint}
\int_0^\infty d \zeta \int_{-\zeta}^{+\zeta} dv \, p(\zeta,v) = 1 \;.
\end{eqnarray}

Fortunately, Eq. (\ref{FKPP_subcritical3}) can be integrated with respect to $v$ upon multiplying by a factor $2 \frac{\partial \mathcal{R}(\zeta,v)}{\partial v}$, yielding
\begin{equation}
\left(\frac{\partial \mathcal{R}(\zeta,v)}{\partial v}\right)^2 = 
\frac{\Delta}{4}\,\mathcal{R}^2(\zeta,v) + \frac{1}{6}\, \mathcal{R}^3(\zeta,v) + \kappa(\zeta),
\label{first_step_subcritical}
\end{equation}
where $\kappa(\zeta)$ is a yet unknown integration constant. To fix $\kappa(\zeta)$, we use the condition in Eq. (\ref{condition}) and arrive at
\begin{eqnarray}
\nonumber
\left(\frac{\partial \mathcal{R}(\zeta,v)}{\partial v}\right)^2 &=& \frac{\Delta}{4}\,
\left( \mathcal{R}^2(\zeta,v) - \mathcal{R}^2(\zeta,0)\right)\\ 
&&+ \frac{1}{6}\, \left( \mathcal{R}^3(\zeta,v) - \mathcal{R}^3(\zeta,0)\right) \;.
\label{second_step_subcritical}
\end{eqnarray}

Since the solution $\mathcal{R}(\zeta,v)$, is symmetric about the $v =0$ line, it is sufficient to solve 
Eq. (\ref{second_step_subcritical}) for only the region $v \in [0,+\zeta]$ (or alternatively for $v \in [-\zeta,0]$).
We restrict on $v \in [0,+\zeta]$ where $\partial {\cal R}(\zeta,v)/\partial v > 0$
(see Fig. \ref{Rzetav_fig}). 
Taking the square root of Eq.~(\ref{second_step_subcritical}) and integrating, we obtain for $v \in [0,+\zeta]$:
\begin{equation}
\int_{\mathcal{R}(\zeta,0)}^{\mathcal{R}(\zeta,v)}  \frac{d r}{\sqrt{\frac{\Delta}{4} \left( r^2 - 
\mathcal{R}^2(\zeta,0)\right) 
+ \frac{1}{6} \left(r^3 - \mathcal{R}^3(\zeta,0)\right)}}
= v.
\label{integral_eqn_subcritical}
\end{equation}
\begin{figure}[ht]
\begin{center}
\includegraphics[width=1\linewidth]{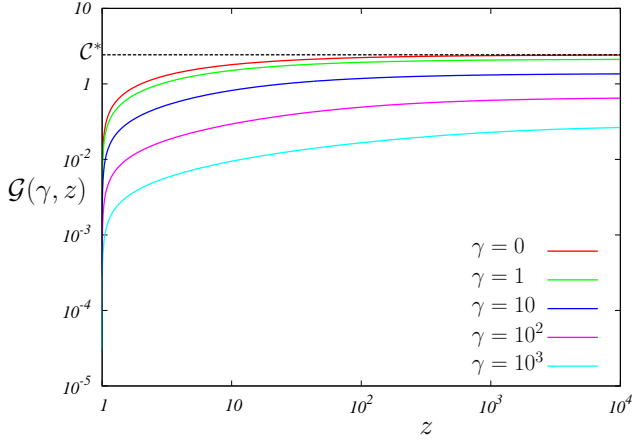}
\end{center}
\caption{The function $\mathcal{G}(\gamma,z)$ for different values of $\gamma$. For large $z$, $\mathcal{G}(\gamma,z)$ saturates
to a $\gamma$ dependent constant $\mathcal{C}(\gamma)$. The case $\gamma = 0$ corresponds to the function ${\mathcal G}(0,z)$ analyzed
in the critical case. The limiting behaviors are ${\mathcal G}(0,z) \to 0$ as $z \to 1$ and ${\mathcal G}(0,z) \to \mathcal{C}^*= 
\frac{\sqrt{\pi}}{3} \frac{\Gamma(\frac{1}{6})}{\Gamma(\frac{2}{3})}
\approx 2.4286$ as $z \to \infty$. }
\label{G_function_figure}
\end{figure}
This equation can be conveniently expressed as
\begin{equation}
\frac{1}{\sqrt{\mathcal{R}(\zeta,0)}} 
\mathcal{G}\left(\frac{3 \Delta/2}{ \mathcal{R}(\zeta,0)},\frac{\mathcal{R}(\zeta,v)}{\mathcal{R}(\zeta,0)}\right) = \frac{v}{\sqrt{6}},
\label{Rsv_subcritical}
\end{equation}
where the bivariate function $\mathcal{G}$ is defined by the integral
\begin{equation}
\mathcal{G}(\gamma,z) = \int_{1}^{z}\frac{dx}{\sqrt{(x^3 -1)+\gamma \left(x^2-1\right)}} \;.
\label{Gdef}
\end{equation}
The above function $\mathcal{G}(\gamma,z)$ can then be expressed as (using the identity 3.138 of Ref. \cite{gradshteyn}):
\begin{eqnarray}
\nonumber
&&\mathcal{G}(\gamma,z) = \\
\nonumber
&&\hspace{-0.5cm}\frac{1}{(3 + 2 \gamma)^{1/4}} {\bf F}\left[2\,\tan^{-1}\sqrt{\frac{z - 1}{\sqrt{3 + 2 \gamma}}}~,\frac{1}{2}\sqrt{2-\frac{3+\gamma}{3+2\gamma}}\right],\\
\label{elliptic_def}
\end{eqnarray}
where $z \ge 1$, $\gamma \ge 0$ and
\begin{equation}
{\bf F}(\phi , k) = \int_{0}^{\phi} \frac{d \theta}{\sqrt{1- k^2 \sin^2 \theta}}
\end{equation}
is the elliptic integral of the first kind. In Fig. \ref{G_function_figure}, we plot the function $\mathcal{G}(\gamma,z)$ as a function of $z$ for different values of $\gamma$. 
Next, inserting the boundary condition $\mathcal{R}(\zeta,\pm \zeta) = 1$ in Eq. (\ref{Rsv_subcritical}) we have
\begin{equation}
\frac{1}{\sqrt{\mathcal{R}(\zeta,0)}} 
\mathcal{G}\left(\frac{3 \Delta/2}{\mathcal{R}(\zeta,0)},\frac{1}{\mathcal{R}(\zeta,0)}\right) = \frac{\zeta}{\sqrt{6}}.
\label{Rstar_subcritical}
\end{equation}
This is an implicit equation for $\mathcal{R}(\zeta,0)$, the solution of which can then be injected in Eq. (\ref{Rsv_subcritical}) to solve 
for $\mathcal{R}(\zeta,v)$ for all $\zeta$ and $v$.

\vspace*{0.5cm}

\noindent{\bf Critical Point.} The computations become slightly more explicit 
exactly at the critical point $a=b$, i.e., $\Delta = 0$. In this case, putting $\gamma = (3 \Delta/2)/{\mathcal R}(\zeta,0) = 0$ in Eq. (\ref{Rstar_subcritical}) gives
\begin{equation}
 \frac{1}{\sqrt{\mathcal{R}(\zeta,0)}} {\mathcal G}\left(0,\frac{\mathcal{R}(\zeta, v)}{\mathcal{R}(\zeta,0)}\right) = \frac{v}{\sqrt{6}} \;,
\label{Rsv_equation_critical}
\end{equation}
where, from Eq. (\ref{Gdef}), ${\mathcal G}(0,z)$ is given by 
\begin{eqnarray}
\nonumber
{\mathcal G}(0,z) &=& \int_{1}^{z} \frac{dx}{\sqrt{x^3 - 1}}\\ 
&=& \frac{2 \sqrt{\pi}\, \Gamma\left( \frac{7}{6}\right)}{\Gamma\left( \frac{2}{3}\right)} - 
\frac{2}{\sqrt{z}} {_2}F_1\left(\frac{1}{6},\frac{1}{2};\frac{7}{6};\frac{1}{z^3}\right),
\label{Y_def}
\end{eqnarray}
where ${_2}F_1$ is the usual Gauss hypergeometric function and $z\ge 1$. This function has the following asymptotic behaviors 
\begin{equation}
{\mathcal G}(0,z) \sim 
\begin{cases}
\dfrac{2}{\sqrt{3}} \sqrt{z -1}~~~~~~~~~~~~\textmd{for $z \rightarrow 1$},\\
\\
\dfrac{\sqrt{\pi}}{3} \dfrac{\Gamma(\frac{1}{6})}{\Gamma(\frac{2}{3})} - \dfrac{2}{\sqrt{z}}~~~~~~\textmd{for $z \rightarrow \infty$}\;.
\end{cases}
\label{y_behaviour}
\end{equation}
Similarly, putting $\Delta = 0$ in Eq. (\ref{Rstar_subcritical}) determines ${\mathcal R}(\zeta,0)$ implicitly as the solution of 
\begin{equation}
\frac{1}{\sqrt{\mathcal{R}(\zeta,0)}} {\mathcal G}\left(0,\frac{1}{\mathcal{R}(\zeta,0)}\right) = \frac{\zeta}{\sqrt{6}} .
\label{Rstar_equation_critical}
\end{equation}
Dividing Eq. (\ref{Rsv_equation_critical}) by (\ref{Rstar_equation_critical}) gives
\begin{equation}
{\mathcal G}\left(0,\frac{\mathcal{R}(\zeta,v)}{\mathcal{R}(\zeta,0)}\right) = \frac{v}{\zeta} {\mathcal G}\left(0,\frac{1}{\mathcal{R}(\zeta,0)}\right) \;,
\label{R_equation}
\end{equation}
where ${\mathcal G}(0,z)$ is given explicitly in Eq. (\ref{Y_def}). The solution of Eq. (\ref{R_equation}) 
thus determines the cumulative joint distribution $\mathcal{R}(\zeta,v)$
in the critical regime. 

\section{Stationary Distribution of the Span}\label{section:span}

The solution $\mathcal{R}(\zeta,v)$ obtained from Eq. (\ref{Rsv_subcritical}) can next be used to compute the stationary joint PDF $p(\zeta,v)$ from Eq. (\ref{joint_PDF_sv}). 
The stationary PDF of the dimensionless span 
$\zeta$, denoted by $p(\zeta)$, is then obtained, for all $\Delta \geq 0$, by integrating over $v$ as:
\begin{equation}\label{margin_zeta}
p(\zeta) = \int_{-\zeta}^{+\zeta} p(\zeta,v) dv \;.
\end{equation} 

%
\begin{figure}[ht]
\begin{center}
\includegraphics[width=\linewidth]{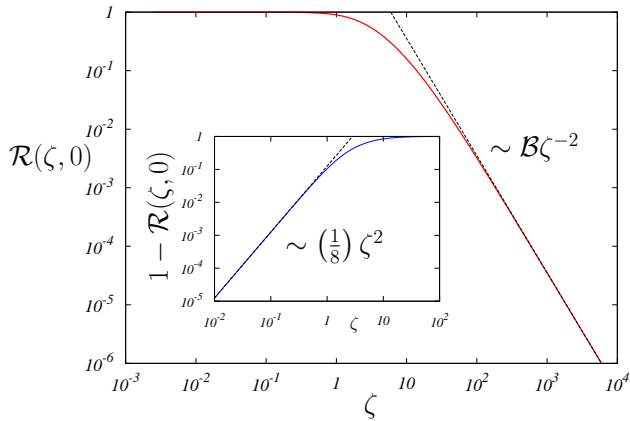}
\end{center}
\caption{The function $\mathcal{R}(\zeta,0)$ versus $\zeta$ in the critical regime derived using Eq. (\ref{Rstar_equation_critical}), 
showing the limiting behaviors $\mathcal{R}(\zeta,0) \to 1$ as $\zeta \to 0$ and 
$\mathcal{R}(\zeta,0) \to \frac{\mathcal{B}}{\zeta^2}$ as $\zeta \to \infty$ (dashed line) as predicted in Eq. (\ref{large_R_behaviour}).
$\mathcal{B} \approx 35.3901$ is defined in Eq. (\ref{constant_definitions}).
{\bf Inset}: Plot of $1 - \mathcal{R}(\zeta,0)$ showing the limiting behavior 
$1 -\mathcal{R}(\zeta,0) \sim \left(\frac{1}{8}\right) \zeta^2$ as $\zeta \to 0$ (dashed line) as predicted in Eq. (\ref{small_R_behaviour2}).}
\label{Rstar_figure}
\end{figure}

%
It is then easy to extract $p(\zeta)$, for all $\Delta \geq 0$, numerically exactly from Eqs.~(\ref{Rsv_subcritical}), (\ref{Rstar_subcritical}), 
(\ref{joint_PDF_sv}) and (\ref{margin_zeta}). 
As an example, we plot this PDF $p(\zeta)$ as a function of $\zeta$ for the critical case ($\Delta = 0$) in Fig. \ref{fig:stationary_dist} and 
for the subcritical case ($\Delta = 1$) in Fig.~\ref{fig:stationary_dist_subcritical}. From the numerical plots for different values of $\Delta$, 
one finds that for small $\zeta$, $p(\zeta)$ increases linearly with a slope that depends on $\Delta$. In contrast, for large $\zeta$, $p(\zeta)$ 
has an algebraic tail $p(\zeta) \propto \zeta^{-3}$ for $\Delta = 0$ while it has an exponential tail for $\Delta > 0$. 
In the next two subsections, we show that these asymptotic behaviors of $p(\zeta)$, both for small and large $\zeta$, can actually be extracted 
analytically for all $\Delta \geq 0$.

\begin{figure}[ht]
\begin{center}
\includegraphics[width=\linewidth]{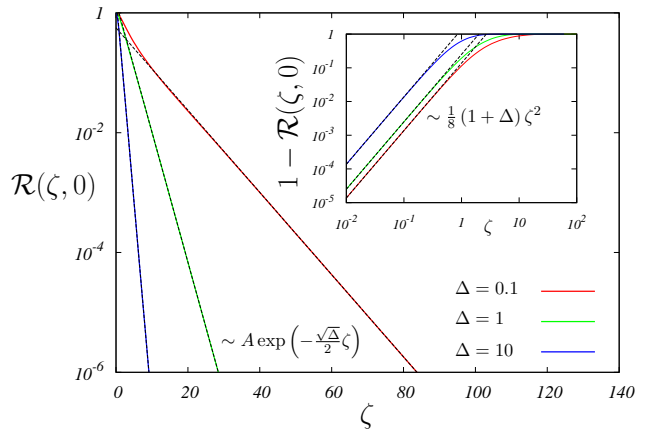}
\end{center}
\caption{Asymptotic behavior of $\mathcal{R}(\zeta,0)$ in the subcritical regime. The plot shows $\mathcal{R}(\zeta,0)$ 
for different values of $\Delta$ derived using Eq. (\ref{Rstar_subcritical}).
The dashed lines representing the asymptotic exponential behavior ${\cal R}(\zeta,0) {\sim} A \exp{\left(-\frac{\sqrt{\Delta}}{2} \zeta\right)}$ as $\zeta \to \infty$ 
derived in Eq. (\ref{large_R_behaviour_subcritical}) are indistinguishable from the theoretically obtained curves as they match exactly. 
{\bf Inset}: Plot of $1 - \mathcal{R}(\zeta,0)$ showing the limiting behavior 
$1-\mathcal{R}(\zeta,0) \sim \frac{1}{8}\left(1 + \Delta \right) \zeta^2$ as $\zeta \to 0$ (dashed lines) as predicted in Eq. (\ref{small_R_behaviour2}).}
\label{Rstar_subcritical_figure}
\end{figure}
\subsection{Asymptotic behavior of $p(\zeta)$ for $\zeta \to 0$}
As $\zeta \to 0$ we have  $\mathcal{R}(\zeta,0) \to 1$. 
Therefore we write 
\begin{equation}
\mathcal{R}(\zeta,0) = 1 - \epsilon(\zeta),  
\end{equation}
where $\epsilon(\zeta)$ is small. Substituting this in Eq. (\ref{Rstar_subcritical}) and expanding ${\cal G}(\gamma,z)$ in Eq. (\ref{Gdef}) around $z = 1$, 
we obtain (using the notation $\epsilon \equiv \epsilon(\zeta)$)
\begin{equation}
\frac{2 \sqrt{\epsilon}}{\sqrt{3(1 + \Delta)}} + \mathcal{O}(\epsilon^{3/2}) =  \frac{\zeta}{\sqrt{6}} .
\end{equation}
Therefore to leading order in $\zeta$ we have
\begin{equation}
\mathcal{R}(\zeta,0) = 1 - \frac{1}{8} (1 + \Delta) \zeta^2 + \mathcal{O}(\zeta^4).
\label{small_R_behaviour2}
\end{equation}
This limiting behavior for the critical case ($\Delta = 0$) is illustrated in the inset of Fig. \ref{Rstar_figure}, and
for the subcritical case in the inset of Fig. \ref{Rstar_subcritical_figure}.
Performing the same analysis in Eq. (\ref{Rsv_subcritical}) with both $\zeta$ and $v$ small gives
\begin{equation}
\mathcal{R}(\zeta,v) = 1 - \frac{1}{8} \left(1 + \Delta \right) \left(\zeta^2 - v^2\right) +  \mathcal{O}(\zeta^4,v^4).
\label{small_R_behaviour3}
\end{equation}
Next, using Eq. (\ref{joint_PDF_sv}), the joint PDF ${p}(\zeta,v)$ is given by
\begin{eqnarray}
p(\zeta,v) &=& \frac{1}{2}\left(\frac{\partial^2}{\partial v^2} -  \frac{\partial^2}{\partial \zeta^2} \right) \mathcal{R}(\zeta,v) \nonumber \\
&=& \frac{1}{4} \left(1 + \Delta \right) +  \mathcal{O}(\zeta^2,v^2).
\end{eqnarray}
Substituting this expression in Eq. (\ref{margin_zeta}) gives 
\begin{equation}
{p}(\zeta) = \frac{1}{2} \left(1 + \Delta \right) \zeta + \mathcal{O}\left(\zeta^3\right) \;,
\label{linear_behaviour_subcritical}
\end{equation}
which yields the small $\zeta$ behavior announced in Eq. (\ref{pzeta_asympt_critical}), for $\Delta = 0$, 
and in Eq. (\ref{pzeta_asympt_subcritical}) for $\Delta > 0$. 
The asymptotic linear growth for small $\zeta$ is shown, for the critical case ($\Delta = 0$), 
in the inset of Fig. \ref{fig:stationary_dist} and for the subcritical case (for $\Delta = 0.1, 1, 10$) 
in the inset of Fig. \ref{fig:stationary_dist_subcritical} {\bf a.}. 
As discussed in section \ref{section:results}, the amplitude of this linear term in (\ref{linear_behaviour_subcritical}) 
carries the signature of the correlation between $x_{\max}$ and $x_{\min}$.

\subsection{Asymptotic behavior of $p(\zeta)$ for $\zeta \to \infty$}

In this subsection we extract analytically the large $\zeta$ tails of $p(\zeta)$ both for the critical ($\Delta =0$) as well as 
for the subcritical case ($\Delta > 0$). 

\subsubsection{Critical point ($\Delta = 0$)}

We start by analyzing Eq. (\ref{Rstar_equation_critical}) in the limit $\zeta \to \infty$. In this limit ${\cal R}(\zeta,0)$ is small. 
We then need to analyze ${\cal G}(0,z)$ for large $z$. Inserting the asymptotic behavior of $\mathcal{G}(0,z)$ in Eq. (\ref{y_behaviour}) 
into Eq (\ref{Rstar_equation_critical}), we obtain
\begin{equation}
\mathcal{R}(\zeta,0) = \frac{\mathcal{B}}{\zeta^2} + \mathcal{O}\left(\frac{1}{\zeta^4}\right),
\label{large_R_behaviour}
\end{equation}
where 
\begin{equation}
\mathcal{B} =  6~ {\mathcal{C}^*}^2 \approx 35.3901,   
~~~~~~\textmd{with}~~~~~~ 
\mathcal{C}^*= \frac{\sqrt{\pi}}{3} \frac{\Gamma(\frac{1}{6})}{\Gamma(\frac{2}{3})}. 
\label{constant_definitions}
\end{equation}
This asymptotic behavior of $\mathcal{R}(\zeta,0)$  is illustrated in Fig.~\ref{Rstar_figure}. Having thus determined ${\cal R}(\zeta,0)$ for large $\zeta$, 
we now investigate ${\cal R}(\zeta,v)$ for large $\zeta$. Our aim is to extract $p(\zeta)$ for large $\zeta$ from Eqs. (\ref{joint_PDF_sv}) and (\ref{margin_zeta}). 
We note that Eq. (\ref{margin_zeta}) involves an integral over $v$ and this integral is dominated by $v \sim \zeta$. 
Hence we need to investigate ${\cal R}(\zeta,v)$ in the scaling limit $\zeta \to \infty$, $v \to \infty$ but keeping $\zeta/v$ fixed. 

When $\zeta \to \infty$, ${\cal R}(\zeta,0) \to 0$ as in Eq. (\ref{large_R_behaviour}). 
As a result, the argument $1/{\cal R}(\zeta,0)$ of ${\cal G}(0,1/{\cal R}(\zeta,0))$ on the right hand side of Eq. (\ref{R_equation}) goes to $\infty$. 
From Eq. (\ref{y_behaviour}), we see that  
${\cal G}(0,z\to \infty) =  \mathcal{C}^*$ where $\mathcal{C}^*$ is given in Eq. (\ref{constant_definitions}). 
Hence, in the scaling limit $\zeta \to \infty$, $v \to \infty$ keeping $\zeta/v$ fixed, Eq. (\ref{R_equation}) becomes
\begin{equation}\label{asympt_scaling}
 {\cal G}\left(0,\frac{{\cal R}(\zeta,v)}{{\cal R}(\zeta,0)} \right) = \mathcal{C}^* \frac{v}{\zeta} \;.
\end{equation}

Inverting the above Eq. (\ref{asympt_scaling}), we get
\begin{equation}
\frac{\mathcal{R}(\zeta,v)}{\mathcal{R}(\zeta,0)} =
\mathcal{F} \left( \mathcal{C}^* \frac{v}{\zeta} \right),
\end{equation}
where $\mathcal{F}(z)$ is defined as the inverse function of ${\mathcal G}(0,z)$. 
Substituting ${\cal R}(\zeta,0) \sim \mathcal{B}/\zeta^2$ from Eq.~(\ref{large_R_behaviour}) gives the final scaling limit expression 
of the joint cumulative distribution

\begin{equation}
\mathcal{R}(\zeta,v) = \frac{\mathcal{B}}{\zeta^2} \mathcal{F} \left(\mathcal{C}^*\frac{v}{\zeta}\right). 
\end{equation}
Inserting this expression into Eq. (\ref{joint_PDF_sv}) we arrive at the joint PDF
\begin{eqnarray}
\nonumber
{p}(\zeta,v)= -\frac{\mathcal{B}}{2} \Big[ \frac{6}{\zeta^4} \mathcal{F}\left( \mathcal{C}^*\frac{v}{\zeta}\right) + 
6 \mathcal{C}^*\frac{v}{\zeta^5} \mathcal{F}'\left( \mathcal{C}^*\frac{v}{\zeta}\right)\\
+ {\mathcal{C}^*}^2\left( \frac{v^2}{\zeta^6} -
\frac{1}{\zeta^4} \right) \mathcal{F}''\left(\mathcal{C}^*\frac{v}{\zeta} \right) \Big].
\end{eqnarray}

\begin{figure}
\begin{center}
\includegraphics[width=1\linewidth]{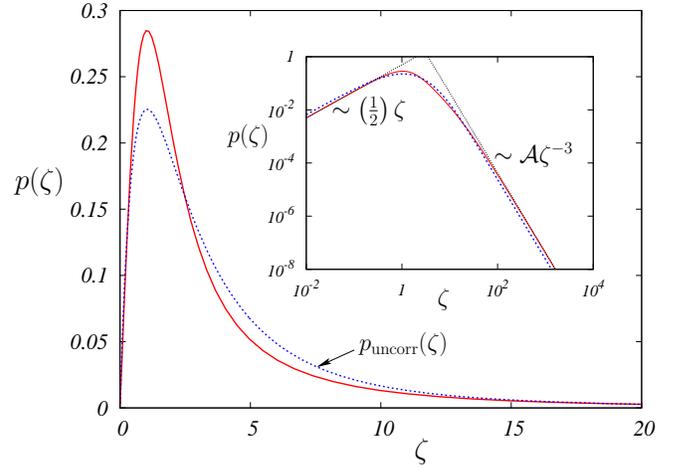}
\end{center}
\caption{Theoretical stationary PDF of the dimensionless span ${p}(\zeta)$ (solid line) in the critical regime ($\Delta = 0$) calculated by 
numerically evaluating Eqs. (\ref{Rstar_equation_critical}), (\ref{R_equation}), (\ref{joint_PDF_sv}) and (\ref{margin_zeta}). 
The (blue) dotted line corresponds to a plot of $p_{\rm uncorr}(\zeta)$ evaluated from Eqs. (\ref{max_distribution_critical}) and (\ref{uncorr_marginal}) 
which is the result for the span distribution obtained
under the assumption that $x_{\max}$ and $x_{\min}$ are uncorrelated. The {\bf Inset} displays the same data in log-log scale 
showing the two limiting behaviors (dashed lines),
linear ${p}(\zeta) \sim \left(\frac{1}{2}\right) \zeta$ for small spans, and 
 ${p}(\zeta) \sim \mathcal{A}/\zeta^3$ for large spans (\ref{pzeta_asympt_critical}), with $\mathcal{A} = 8 \pi \sqrt{3} \simeq 43.53118\ldots$.
}
\label{fig:stationary_dist}
\end{figure}

Substituting this expression for $p(\zeta,v)$ in Eq. (\ref{margin_zeta}) gives finally
%
\begin{eqnarray}
\nonumber
{p}(\zeta) = -\frac{1}{\zeta^3} \left( \frac{\mathcal{B}}{\mathcal{C}^*}\right) \int_{0}^{\mathcal{C}^*} dz  \Big[ 6 \mathcal{F}\left( z\right)
+  6 z \mathcal{F}'\left(z\right) + \\
\left(z^2- {\mathcal{C}^*}^2\right) \mathcal{F}''\left(z\right) \Big].
\end{eqnarray}
Hence, we obtain 
\begin{equation}
{p}(\zeta) \sim  \frac{\mathcal{A}}{\zeta^3}~~~~~~~\textmd{for large $\zeta$},
\end{equation}
with
\begin{equation}
\mathcal{A} =  -6 ~\mathcal{C}^*\int_{0}^{\mathcal{C}^*} dz \left[6 \mathcal{F}(z) 
+ 6 z \mathcal{F}'(z) + (z^2 - {\mathcal{C}^*}^2)\mathcal{F}''(z) \right].
\label{A_def}
\end{equation}
It turns out that the integral in Eq. (\ref{A_def}) can be performed explicitly (see 
Appendix \ref{critical_prefactor_appendix}) and the final answer for the amplitude is amazingly
simple
\begin{equation}\label{amplitude_A}
{\cal A} = 8 \pi \sqrt{3} = 43.53118\ldots \;.
\end{equation}
Thus the leading asymptotic behavior for large $\zeta$ is
\begin{equation}
{p}(\zeta) \sim  \frac{8 \pi \sqrt{3}}{\zeta^3}  \;,
\label{final_power_law}
\end{equation}
as announced in Eq. (\ref{pzeta_asympt_critical}). 
This large $\zeta$ behavior of the span PDF is shown in Fig. \ref{fig:stationary_dist} where we plot the 
numerically exact distribution ${p}(\zeta)$ (extracted from Eqs. (\ref{Rstar_equation_critical}), (\ref{R_equation}), (\ref{joint_PDF_sv}) and (\ref{margin_zeta})),  
along with the asymptotic power law tail 
derived analytically in Eq. (\ref{final_power_law}).
At large $\zeta$, the exact distribution shows a good agreement with the asymptotic behaviour.
We also display $p_{\rm uncorr}(\zeta)$, the span PDF obtained assuming that $x_{\max}$ and $x_{\min}$ are uncorrelated, 
which decays with the same power, but with a different prefactor provided in Eq. (\ref{uncorr_large_critical}).
The amplitude ${\cal A} = 8 \pi \sqrt{3}$ is thus nontrivial due to the remnant non-vanishing 
correlation between $x_{\max}$ and $x_{\min}$ in the stationary state 
(see also the discussion at the end of section \ref{section:results}). 

\begin{figure}
\begin{center}
{\bf a.}
\includegraphics[width=0.9\linewidth]{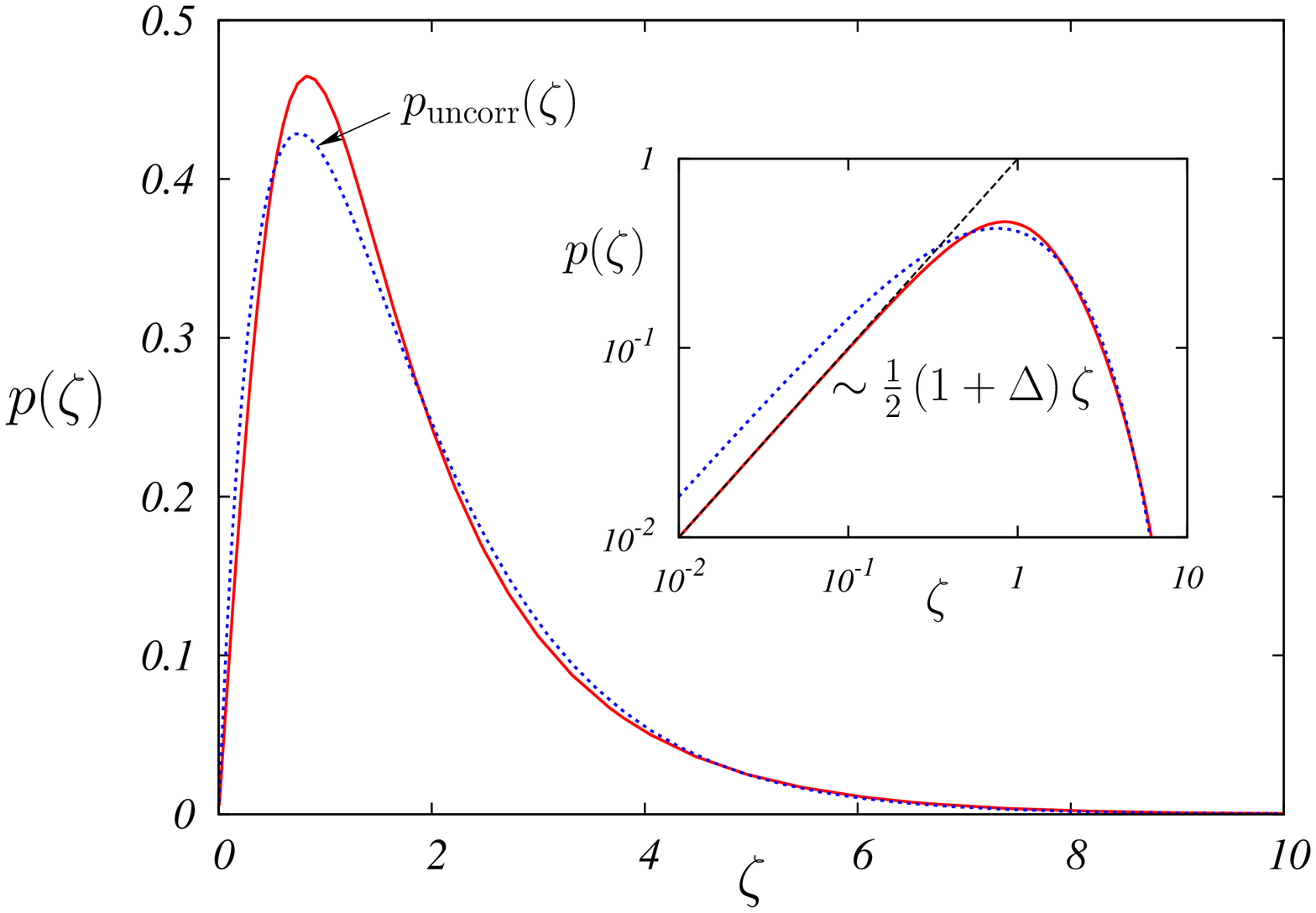}\\
{\bf b.}
\includegraphics[width=0.9\linewidth]{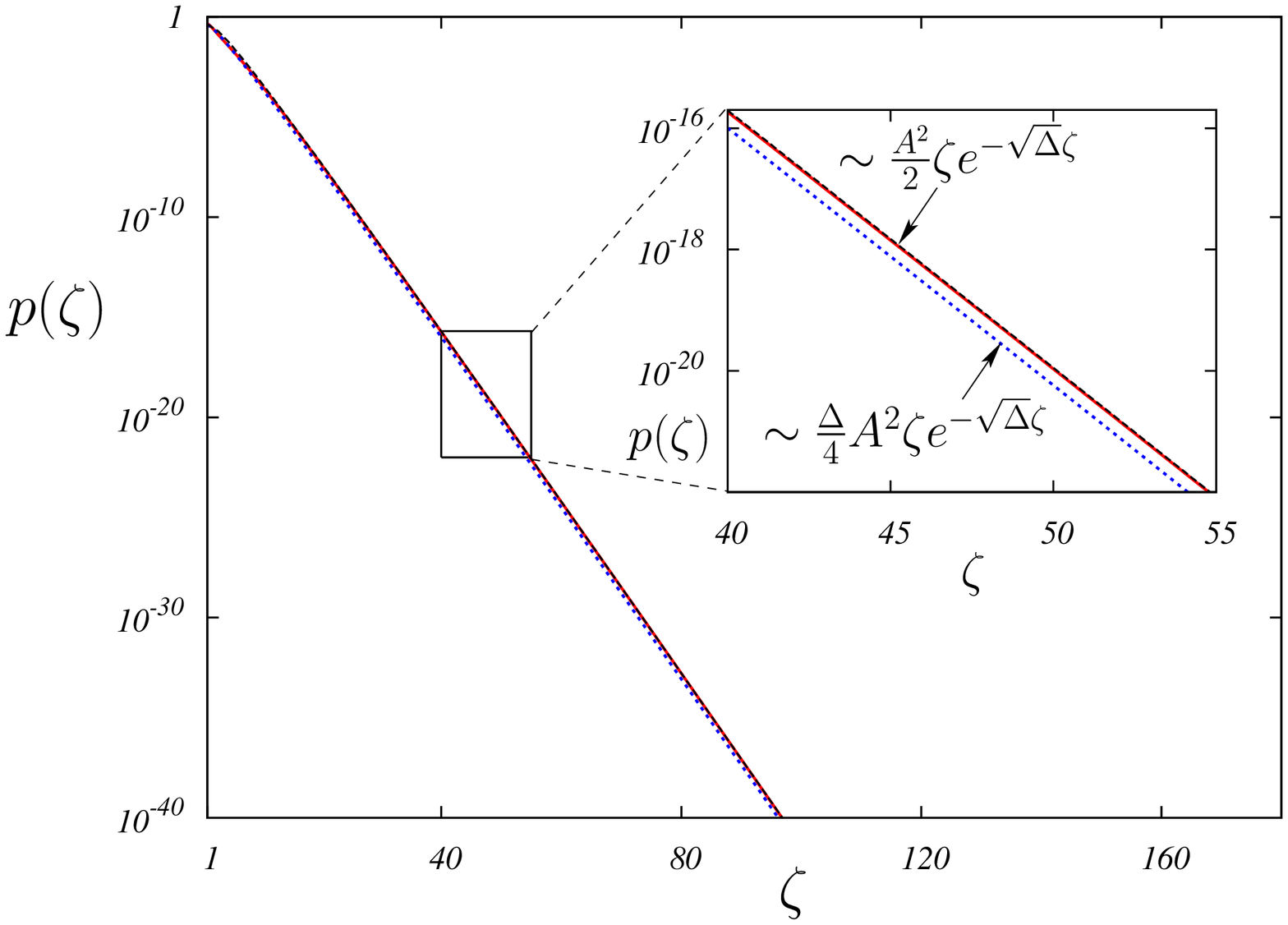}
\end{center}
\caption{Theoretical stationary PDF of the dimensionless span $p(\zeta)$ in the subcritical regime.
In {\bf a.} we display $p(\zeta)$ (solid line), for $\Delta = 1$, 
calculated by numerically evaluating 
Eqs. (\ref{Rsv_subcritical}), (\ref{Rstar_subcritical}), (\ref{joint_PDF_sv}) and (\ref{margin_zeta}). 
The (blue) dotted line corresponds to $p_{\rm uncorr}(\zeta)$ evaluated
from Eqs. (\ref{max_distribution_subcritical}) and (\ref{uncorr_marginal}) which is the result for the span distribution obtained
under the assumption that $x_{\max}$ and $x_{\min}$ are uncorrelated. The {\bf Inset} displays the same data in log-log scale 
showing the approach to the 
$\zeta \to 0$ linear behavior (dashed line) derived in Eq. (\ref{linear_behaviour_subcritical}). 
In {\bf b.} we display the stationary span PDF $p(\zeta)$ and $p_{\rm uncorr}(\zeta)$ for $\Delta =1$ in log-linear scale.
The dashed line showing the $\zeta \to \infty$ 
asymptotic behavior $\sim \frac{A^2}{2} \zeta \exp (- \sqrt{\Delta} ~\zeta)$ derived in Eq. (\ref{large_zeta_p_subcritical}) 
is indistinguishable from the theoretically obtained curve as they match exactly. 
The {\bf Inset} shows the same data highlighting
the difference between the asymptotic decays of $p(\zeta)$ and $p_{\rm uncorr}(\zeta)$ 
(the asymptotic decay of $p_{\rm uncorr}(\zeta)$ is provided in Eq. (\ref{uncorr_large_final})).}
\label{fig:stationary_dist_subcritical}
\end{figure}

\subsubsection{Subcritical case ($\Delta > 0$)}

In the subcritical regime, the large $\zeta$ analysis can be done along the same lines as before for the critical 
case but the analysis is a bit more complicated and most of the details have been relegated to Appendix 
\ref{subcritical_asymptotics_appendix}.

As in the critical case, here also we first need to compute the large $\zeta$ behavior of 
${\cal R}(\zeta,0)$ from Eq. (\ref{Rstar_subcritical}) with $\Delta>0$. This is done in
Appendix \ref{subcritical_asymptotics_appendix}. We find (see Eq. (\ref{expansion_R0}))
\begin{eqnarray}
\nonumber
{\cal R}(\zeta,0) \sim A \exp{\left(-\frac{\sqrt{\Delta}}{2} \zeta\right)} \;, \; {\rm as} \;\; \zeta \to \infty \;\\
{\rm where} \;\; A = \frac{12 \Delta}{\left[ \sqrt{3\Delta/2} + \sqrt{1+3\Delta/2}\right]^2}  \;.\label{large_R_behaviour_subcritical}
\end{eqnarray}
In Fig. \ref{Rstar_subcritical_figure} we show a plot of ${\cal R}(\zeta,0)$ obtained by numerically 
evaluating Eq. (\ref{Rstar_subcritical}) for different values of $\Delta$. 
For large $\zeta$, this shows a good agreement with the asymptotic behavior in Eq. (\ref{large_R_behaviour_subcritical}). 
Using this asymptotic tail of ${\cal R}(\zeta,0)$ from Eq. (\ref{large_R_behaviour_subcritical}),
we then analyse the asymptotic behavior of ${\cal R}(\zeta,v)$ in Eq. (\ref{Rsv_subcritical}), in the scaling limit 
$\zeta \to \infty$, $v \to \infty$, keeping $\zeta/v$ fixed. We find
\begin{equation}\label{general_exp_text}
{\cal R}(\zeta,v) = H_0(\zeta,v) + {\cal R}^2(\zeta,0) H_1(\zeta,v) + {\cal O}\left({\cal R}^3(\zeta,0) \right) \;.
\end{equation}
where both $H_0(\zeta,v)$ and $H_1(\zeta,v)$ are functions of $u = \zeta - v$ only, 
whose expressions are provided in Eq. (\ref{eq_H0_2}) and Eq. (\ref{def_H1}) respectively. From ${\cal R}(\zeta,v)$, 
we can then obtain the joint PDF $p(\zeta,v)$ from Eq. (\ref{joint_PDF_sv}) and eventually $p(\zeta)$ from Eq. (\ref{margin_zeta}). 
Following this procedure, we finally obtain the large $\zeta$ behavior of $p(\zeta)$ for $\Delta > 0$ (see Appendix \ref{subcritical_asymptotics_appendix} for details): 
\begin{equation}\label{large_zeta_p_subcritical}
p(\zeta) = \frac{A^2}{2}\, \zeta\,  e^{-\sqrt{\Delta}~\zeta} \left(1 + {\cal O}(\zeta^{-1}) \right) \;,
\end{equation}
with $A$ given in Eq. (\ref{large_R_behaviour_subcritical}), as announced in Eq. (\ref{pzeta_asympt_subcritical}). 
Here again, as discussed in section \ref{section:results}, the amplitude ${A^2}/{2}$ in Eq. (\ref{large_zeta_p_subcritical}) 
bears the signatures of the correlations between $x_{\max}$ and $x_{\min}$. 
In Fig. \ref{fig:stationary_dist_subcritical} we show a plot of $p(\zeta)$, for the subcritical case (for $\Delta = 1$), 
obtained by numerically evaluating Eqs. (\ref{Rsv_subcritical}), (\ref{Rstar_subcritical}), (\ref{joint_PDF_sv}) and (\ref{margin_zeta}). 
For comparison, we also show a plot of $p_{\rm uncorr}(\zeta)$ obtained from Eqs. (\ref{max_distribution_subcritical}) and (\ref{uncorr_marginal}) 
which corresponds to the PDF of the span obtained by assuming that $x_{\max}$ and $x_{\min}$ are independent. We also display the agreement of the 
asymptotic behaviors for $\zeta \to 0$ and $\zeta \to \infty$ derived in 
Eqs. (\ref{linear_behaviour_subcritical}) and (\ref{large_zeta_p_subcritical}) with these numerically exact PDFs.

\section{Monte Carlo Simulations}

We have performed numerical simulations of the BBM and numerically computed the PDF of the span
at different times $t$. Directly simulating the BBM model
is in general hard to do in the supercritical regime ($b>a$) where there is an exponential proliferation of particles. 
In this case one has to resort to numerically evaluating the non-linear FKPP-type equations to extract the behavior of the PDF at large times
\cite{brunet_derrida_epl,brunet_derrida_jstatphys}. However, 
in the critical ($b=a$) and subcritical ($b < a$) cases, it is possible to obtain very good statistics by performing direct Monte Carlo
simulations of the process \cite{ramola_majumdar_schehr}.

\begin{figure}[hh]
\begin{center}
{\bf a.}\includegraphics[width=0.9\linewidth]{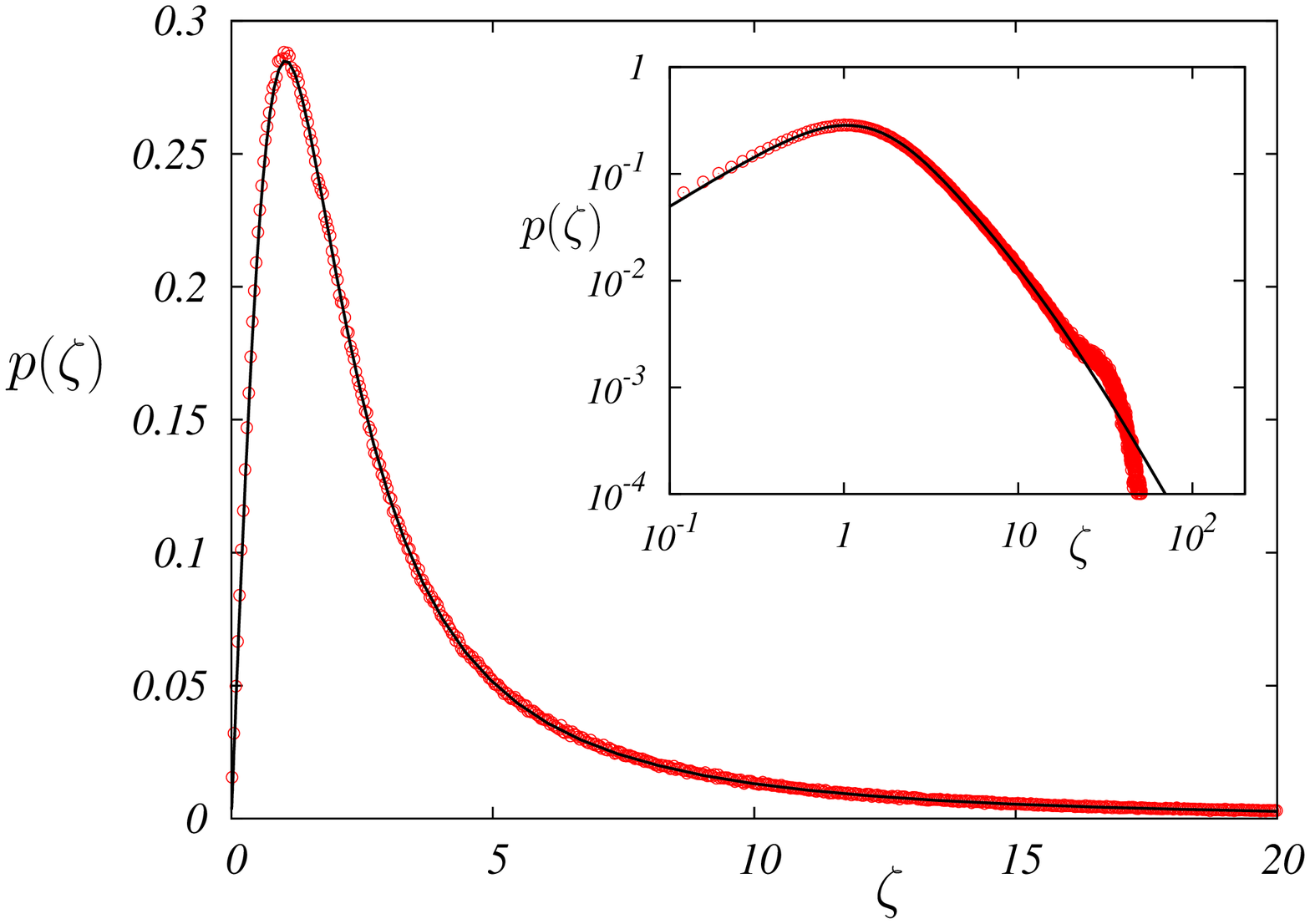}\\
{\bf b.}\includegraphics[width=0.9\linewidth]{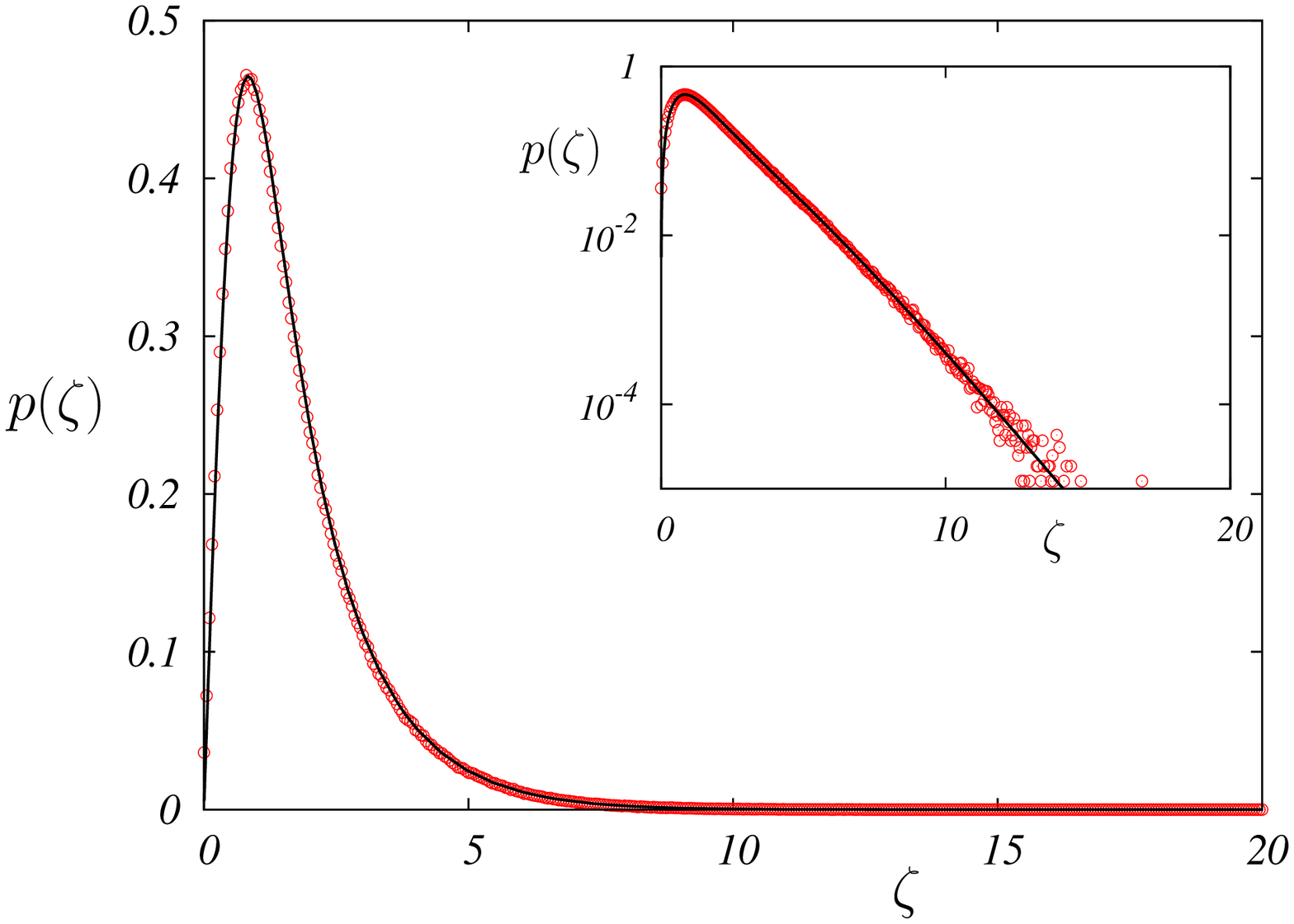}
\end{center}
\caption{{\bf a.} Probability distribution function of the dimensionless span $p(\zeta)$ extracted from Monte Carlo simulations (open circles) in the critical case ($\Delta=0$). 
Here $t = 100$, $D = 1$, $a = b = 1$, and $dt = 0.0001$.
The data is averaged over $5 \times 10^7$ realizations.
The bold line represents the stationary theoretical PDF $p(\zeta)$
extracted numerically from Eqs. (\ref{Rstar_equation_critical}), (\ref{R_equation}), (\ref{joint_PDF_sv}) and (\ref{margin_zeta}).
 The {\bf Inset} displays the same data
over a wider range in log-log scale. The curvature of the PDF for large $\zeta$ arises due to the finite time
nature of our measurement.\\
{\bf b.} 
Probability distribution function of the dimensionless span $p(\zeta)$ extracted from Monte Carlo simulations (open circles) in the subcritical regime. 
Here $t = 100$, $D = 1$, $a = 2$, $b = 1$ (i.e. $\Delta = 1$), and $dt = 0.0001$.
The data is averaged over $5 \times 10^7$ realizations.
The bold line represents the stationary theoretical PDF $p(\zeta)$
extracted numerically from Eqs. (\ref{Rsv_subcritical}), (\ref{Rstar_subcritical}), (\ref{joint_PDF_sv}) and (\ref{margin_zeta}).
The {\bf Inset} displays the same data
in log-linear scale showing the approach to the exponential behaviour for large $\zeta$ as predicted in Eq. (\ref{large_zeta_p_subcritical}).
}
\label{fig:MonteCarlo}
\end{figure}

In Fig. \ref{fig:MonteCarlo} {\bf a.} we show our numerical results at criticality ($\Delta = 0$) for the PDF of the 
dimensionless span 
$P(\zeta,\tau = b t)$ at large time $t = 100$ (note that the discrete time step was set to 
$\Delta t = 0.0001$ in our simulations) with $D =1$ and $a = b = 1$. These data show a very good agreement with our 
theoretical predictions for the stationary PDF $p(\zeta)$  
extracted numerically from Eqs. (\ref{Rstar_equation_critical}), (\ref{R_equation}), (\ref{joint_PDF_sv}) and (\ref{margin_zeta}). 
In Fig.~\ref{fig:MonteCarlo} {\bf b.} we show the same quantity in the subcritical regime for 
$\Delta = 1$. Here again, we observe a very good agreement between the Monte Carlo simulations and our exact 
theoretical results extracted numerically from Eqs. (\ref{Rsv_subcritical}), (\ref{Rstar_subcritical}), (\ref{joint_PDF_sv}) and (\ref{margin_zeta}),
 except in the small $\zeta$ region where discretization effects become important. 
We have checked that as we decrease the size of the time steps, the results from our simulations match more closely with our theoretical predictions. 

Finally, we have also numerically studied the finite time behavior of the solution of Eq.~(\ref{FKPP_R}). In particular, for $\Delta = 0$, 
our data for finite times indicate that $P(\zeta,\tau)$ takes the scaling form $P(\zeta,\tau) \sim p(\zeta) {\cal S}(\zeta/\sqrt{\tau})$, 
where ${\cal S}(u)$ is a rapidly decaying function as $u \to \infty$. The curvature of the PDF at large $\zeta$ in the inset of Fig. \ref{fig:MonteCarlo} {\bf a.} 
is a feature that emerges due to the finite time nature of our measurement, and our numerical solutions reproduce it exactly.
It would be interesting to analytically study this finite time behavior
which certainly deserves further investigation. 

\section{Conclusion}

In summary, we have obtained exact results for the stationary joint PDF $p(x,y)$ of the dimensionless maximal displacements 
$x_{\max} = X_{\max}/\sqrt{D/b}$ and $x_{\min} = X_{\min}/\sqrt{D/b}$ {\it up to time $t$} for the one-dimensional BBM in the critical ($a=b$) 
and subcritical cases ($a>b$) (see Figs. \ref{walk_picture} and \ref{fig:cartoon}). 
In both cases we found that the correlation between $x_{\max}$  and $x_{\min}$ remain nonzero, even in the stationary state. 
From this joint PDF we have computed exactly the PDF $p(\zeta)$ of the (dimensionless) span, $\zeta = (X_{\max} - X_{\min})/\sqrt{D/b}$, 
which provides an estimate of the spatial extent of the process. 
We demonstrated that $p(\zeta)$ carries the signatures of the correlation between the two extreme displacements $x_{\max}$ and $x_{\min}$, 
which can be seen for instance in the asymptotic behaviors of $p(\zeta)$ both for small and large arguments (\ref{pzeta_asympt_critical}, \ref{pzeta_asympt_subcritical}).  

The span is an interesting physical observable associated with BBM, which has several potential applications, for example in the context of epidemic spreads \cite{majumdar_pnas}. 
Moreover, our results are also interesting from the general point of view of extreme value statistics (EVS) of strongly correlated variables.
It was indeed recently demonstrated that random walks and Brownian motion 
(see e.g., Refs. \cite{brunet_derrida_epl,brunet_derrida_jstatphys,kundu,satya_airy1,satya_airy2,schehr_majumdar,perret,mounaix} for recent studies) 
are interesting laboratories to test the effects of correlations on EVS, 
beyond the well known case of independent and identical random variables \cite{gumbel}. 
In that respect, the results for the one-dimensional BBM obtained in the present paper constitute an interesting instance of a strongly correlated multi-particle 
system where the correlation between extreme values can be computed analytically.  

In this paper we have restricted ourselves to computing the span distribution for the critical ($b=a$) and subcritical cases ($b<a$). The computation was feasible because the span distribution becomes stationary at late times $t$ in these cases. In contrast, in the supercritical case ($b>a$) the
span distribution will always be time dependent and it would be interesting to compute this distribution exactly. The recent developments \cite{brunet_derrida_epl,brunet_derrida_jstatphys,ABK12,ABBS13,ABK13} in the supercritical case may shed some light on this outstanding problem.

It would also be interesting to extend these calculations to branching processes where the particles can split into $k>2$ 
particles at each time step, which can be treated using the techniques developed in our paper.

\section*{Acknowledgements}
K.~R. acknowledges helpful discussions with A. Kundu, A. Gudyma, C. Texier and B. Derrida. 
SNM and GS acknowledge support by ANR grant
2011-BS04-013-01 WALKMAT and in part by the Indo-
French Centre for the Promotion of Advanced Research
under Project 4604-3. 




\appendix
\section{Critical Prefactor}
\label{critical_prefactor_appendix}
In this appendix we derive an exact and simple expression for the amplitude $\mathcal{A}$ of the power law decay of the span PDF
in the critical regime, given in Eq. (\ref{A_def}) in the text. We first rewrite the integral in Eq. (\ref{A_def}) as
\begin{eqnarray}
\nonumber
&&\mathcal{A} = \\
\nonumber
&&\hspace{-0.5cm}-6 ~\mathcal{C}^*\int_{0}^{\mathcal{C}^*} dz \left\{2 \mathcal{F}(z) 
+ \left[ 4 z \mathcal{F}(z) + (z^2 - {\mathcal{C}^*}^2)\mathcal{F}'(z) \right]'\right\}.\\
\label{A_def2}
\end{eqnarray}
Integrating with respect to $z$ we arrive at
\begin{eqnarray}
\nonumber
\mathcal{A} = -6 ~\mathcal{C}^* \left[ 4 z \mathcal{F}(z) + (z^2 - {\mathcal{C}^*}^2)\mathcal{F}'(z) \right] \Big\lvert_{0}^{\mathcal{C}^*}\\
-12  ~\mathcal{C}^*\int_{0}^{\mathcal{C}^*} dz \mathcal{F}(z).
\label{A_def3}
\end{eqnarray}
Next, it is convenient to perform the change of variable $r = \mathcal{F}(z)$, with $z = {\mathcal{G}}(0,r)$ and $r \in [1,\infty]$
when $z \in [0, \mathcal{C}^*]$, where the function ${\mathcal{G}}(0,r)$ is defined in Eq. (\ref{Y_def}).
$\mathcal{F}(z)$ and its derivative can then be expressed as follows
\begin{eqnarray}
\nonumber
&&\mathcal{F}(z) =  r,\\
&&\mathcal{F}'(z) = \frac{1}{{\frac{d}{dr}\mathcal{G}}(0,r)} = \sqrt{r^3 -1}.
\end{eqnarray}
Similarly we can represent the integral in Eq. (\ref{A_def3}) in terms of the following function.
We define  
\begin{eqnarray}
\nonumber
&&\int_z^{\mathcal{C}^*} dz' \, \mathcal{F}(z') \equiv \mathcal{H}(r) = \int_r^{\infty} dr' \frac{r'}{\sqrt{{r'}^{3}-1}}\\
&&= -\frac{2 \sqrt{\pi} \Gamma\left( \frac{5}{6}\right)}{\Gamma\left( \frac{1}{3}\right)} + 
2 \sqrt{r} ~{_2}F_1\left(\frac{1}{6},\frac{1}{2};\frac{7}{6};\frac{1}{r^3}\right).
\label{H_def}
\end{eqnarray}
Inserting the above expressions into Eq. (\ref{A_def3}), we arrive at the following exact expression for the coefficient
\begin{equation}
\mathcal{A}  = \mathcal{I}(r)\Big\lvert_{1}^{\mathcal{\infty}},
\label{A_def4}
\end{equation}
where the integrand $\mathcal{I}(r)$ is defined as
\begin{eqnarray}
\nonumber
&&\mathcal{I}(r) =\\
\nonumber
&&\hspace*{-0.5cm}-6 ~\mathcal{C}^*\left[ 2 \mathcal{H}(r) + 4 ~r {\mathcal{G}(0,r)} + 
\left({\mathcal{G}(0,r)}^2 - {\mathcal{C}^*}^2\right) \sqrt{r^3 - 1} \right]\;.\\ 
\label{def_I}
\end{eqnarray}
We next examine the limiting behaviors of the integrand $\mathcal{I}(r)$ in (\ref{def_I}). 
Using the expressions of the functions
${\mathcal{G}}(0,r)$ in Eq. (\ref{Y_def}) and $\mathcal{H}(r)$ in Eq. (\ref{H_def}), we find
\begin{flalign}
\mathcal{I}(r) \sim 
\begin{cases}
\left(48 \sqrt{3 \pi } \frac{\Gamma \left(\frac{7}{6}\right) \left(\pi  \Gamma \left(\frac{7}{6}\right)^2-\Gamma
\left(\frac{2}{3}\right)^2\right)}{\Gamma \left(\frac{2}{3}\right)^3}\right) \sqrt{r-1} \;,r \to 1,\\
8 \pi \sqrt{3}  - \left(\frac{288 \pi  \Gamma \left(\frac{7}{6}\right)^2}{7 \Gamma \left(\frac{2}{3}\right)^2} \right)  \frac{1}{r^2} \;,r \to \infty.
\end{cases}
\end{flalign}
Using the above expressions we obtain the limiting behaviors 
\begin{equation}
\nonumber
\lim_{r \to 1} \mathcal{I}(r) \to 0 ~~~~~~~~\textmd{and}~~~~~~~~
\lim_{r \to \infty} \mathcal{I}(r) \to 8 \pi \sqrt{3} \;. 
\end{equation}
Finally, inserting these into Eq. (\ref{A_def4}), we obtain the following exact value for
the coefficient
\begin{equation}
\mathcal{A} =  8 \pi \sqrt{3} \simeq 43.53118\ldots,
\end{equation}
as announced in the text (\ref{amplitude_A}).

\section{Asymptotic behaviors of the PDF of the span in the subcritical case $\Delta >0$}
\label{subcritical_asymptotics_appendix}


We recall that the PDF of the span $p(\zeta)$ is given by Eq. (\ref{joint_PDF_sv}):
\begin{equation}\label{pzeta_app}
p(\zeta) = \int_{-\zeta}^{+\zeta} p(\zeta,v) dv = 2\int_0^{+\zeta} p(\zeta,v) dv \;,
\end{equation}
where we have used $p(\zeta,v) = p(\zeta,-v)$ and where $p(\zeta,v)$ is given by 
\begin{equation}
{p}(\zeta,v)  =  -\frac{1}{2}\frac{\partial}{\partial x} \frac{\partial}{\partial y} \mathcal{R}(x,y) \equiv \frac{1}{2} 
\left(\frac{\partial^2}{\partial v^2} -  \frac{\partial^2}{\partial \zeta^2} \right) \mathcal{R}(\zeta,v).
\label{joint_PDF_sv_app}
\end{equation}
The function ${\cal R}(\zeta,v)$ is itself determined implicitly by Eq. (\ref{Rsv_subcritical})
\begin{equation}
\frac{1}{\sqrt{\mathcal{R}(\zeta,0)}} 
\mathcal{G}\left(\frac{3 \Delta/2}{ \mathcal{R}(\zeta,0)},\frac{\mathcal{R}(\zeta,v)}{\mathcal{R}(\zeta,0)}\right) = \frac{v}{\sqrt{6}},
\label{Rsv_subcritical_app}
\end{equation}
where $\mathcal{G}(\gamma,z)$ is given by
\begin{equation}
\mathcal{G}(\gamma,z) = \int_{1}^{z}\frac{dx}{\sqrt{(x^3 -1)+\gamma \left(x^2-1\right)}} \;.
\label{Gdef_app}
\end{equation}
The goal is to extract the behavior of the function ${\cal R}(\zeta,v)$ from Eqs. (\ref{Rsv_subcritical_app}), (\ref{Gdef_app}) 
in the limit of large $\zeta$ and $v$ with $v \sim \zeta$ -- as the integral over $v$ in Eq. (\ref{pzeta_app}) is dominated by $v \sim \zeta$.

First, it is convenient to rewrite the function $\mathcal{G}(\gamma,z)$ in (\ref{Gdef_app}) as
\begin{equation}\label{decomp_G}
\mathcal{G}(\gamma, z) =  \mathcal{G}_{\infty}(\gamma) - \widetilde{\mathcal{G}}(\gamma,z)
\end{equation}
where
\begin{flalign}
\nonumber
\mathcal{G}_{\infty}(\gamma) = \int_{1}^{\infty}\frac{dx}{\sqrt{(x^3 -1)+\gamma \left(x^2-1\right)}},\\ 
\widetilde{\mathcal{G}}(\gamma,z)=\int_{z}^{\infty}\frac{dx}{\sqrt{(x^3 -1)+\gamma \left(x^2-1\right)}}.
\label{gfunction_expansion}
\end{flalign}
In the following, we will need the asymptotic behavior of $\mathcal{G}_{\infty}(\gamma)$ for large $\gamma$: 
\begin{equation}\label{ginf_largegamma}
\mathcal{G}_{\infty}(\gamma) = \frac{1}{\sqrt{\gamma}} \left(\log \gamma + 3 \log 2 \right) - \frac{1}{2 \gamma^{3/2}}+ \mathcal{O}\left(\frac{1}{\gamma^{5/2}}\right) \;.
\end{equation}
On the other hand, we also need the asymptotic expansion of the function $\widetilde{\mathcal{G}}(\gamma,z)$ 
in (\ref{gfunction_expansion}) in the limit $\gamma \to \infty$, $z \to \infty$, keeping the ratio $\gamma/z = \alpha$ fixed. 
This expansion can be obtained straightforwardly from the integral representation given in (\ref{gfunction_expansion}) by performing the change of variable $x = zu$. One finds
\begin{equation}
\widetilde{\mathcal{G}}(\gamma,z) = \frac{1}{\sqrt{z}} \left( \widetilde{\mathcal{G}_{1}}\left(\frac{\gamma}{z}\right) +  
\frac{1}{z^{2}}\widetilde{\mathcal{G}_{2}}\left(\frac{\gamma}{z}\right) + {\cal O}\left(\frac{1}{z^{3}}\right)\right)
\label{gtilde_expansion}
\end{equation}
where
\begin{eqnarray}
\label{def_G1}
\widetilde{\mathcal{G}_{1}}(\alpha) =  
\int_{1}^{\infty}\frac{du}{\sqrt{\left(u^3 + \alpha u^2\right)}} &=& \frac{2 \sinh ^{-1}(\sqrt{\alpha})}{\sqrt{\alpha}} \\
&{\sim}& \frac{\ln \alpha + \ln 4}{\sqrt{\alpha}} \;, \; {\rm as} \; \alpha \to \infty \nonumber
\end{eqnarray}
and 
\begin{eqnarray}
\label{def_G2}
\nonumber
&&\widetilde{\mathcal{G}_{2}}(\alpha) = 
\frac{\alpha}{2} \,\int_{1}^{\infty} \frac{ du}{\left(u^3 +\alpha u^2\right)^{3/2}}\\
&&= \frac{\sqrt{\alpha } \left(2 \alpha ^2-5 \alpha -15\right)+15 \sqrt{\alpha +1} \sinh ^{-1}\left(\sqrt{\alpha }\right)}{8
   \sqrt{\alpha ^5 (\alpha +1)}}  \nonumber \\
&&{\sim} \frac{1}{4 \sqrt{\alpha}} \;, \; {\rm as}\; \; \alpha \to \infty \;.
\end{eqnarray}

Solving Eq. (\ref{Rsv_subcritical_app}) for ${\cal R}(\zeta,v)$ requires the knowledge of ${\cal R}(\zeta,0)$ which we first study, in the large $\zeta$ limit. 
This function ${\cal R}(\zeta,0)$ satisfies (as given in Eq. (\ref{Rstar_subcritical}))
\begin{equation}
\frac{1}{\sqrt{\mathcal{R}(\zeta,0)}} 
\mathcal{G}\left(\frac{3 \Delta/2}{\mathcal{R}(\zeta,0)},\frac{1}{\mathcal{R}(\zeta,0)}\right) = \frac{\zeta}{\sqrt{6}}.
\label{Rstar_subcritical_app}
\end{equation}
In the limit $\zeta \to \infty$, one expects that ${\mathcal R}(\zeta,0) \to 0$ hence we need the asymptotic behavior of ${\cal G}(\gamma,z)$ 
for $\gamma = {(3 \Delta/2)}/{\mathcal{R}(\zeta,0)} \to \infty$, $z = {1}/{\mathcal{R}(\zeta,0)} \to \infty$ keeping $\alpha = \gamma/z = 3\Delta/2$ fixed. 
Using Eq. (\ref{decomp_G}) together with the asymptotic expansion in Eq. (\ref{gtilde_expansion}) at lowest order -- i.e, retaining only $\widetilde{{\cal G}_1}(\alpha)$ -- one 
finds from Eq. (\ref{Rstar_subcritical_app})
\begin{eqnarray}
\label{expansion_R0}
\nonumber
{\cal R}(\zeta,0) {\sim} A \exp{\left(-\frac{\sqrt{\Delta}}{2} \zeta\right)} \;, \; {\rm as} \;\; \zeta \to \infty \\
{\rm where} \;\; A = \frac{12 \Delta}{\left[ \sqrt{3\Delta/2} + \sqrt{1+3\Delta/2}\right]^2} \;,
\end{eqnarray}
as given in Eq. (\ref{large_R_behaviour_subcritical}) in the text. 

We now study the asymptotic expansion of ${\cal R}(\zeta,v)$, from Eq. (\ref{Rsv_subcritical_app}) and using the expansion of ${\cal R}(\zeta,0)$ obtained above (\ref{expansion_R0}). 
Here we need the asymptotic behavior of ${\cal G}(\gamma,z)$ for 
\begin{equation}
\gamma = {(3 \Delta/2)}/{\mathcal{R}(\zeta,0)} \to \infty \;, \; z = {{\cal R}(\zeta,v)}/{\mathcal{R}(\zeta,0)} \to \infty \nonumber
\end{equation}
keeping $\alpha = \gamma/z = 3\Delta/(2{\cal R}(\zeta,v)) $ fixed. 
Inserting the asymptotic expansions obtained above (\ref{ginf_largegamma})-(\ref{gtilde_expansion}) in Eq. (\ref{Rsv_subcritical_app}) one finds
\begin{eqnarray}
\label{explicit_expansion}
\nonumber
&&\frac{1}{\sqrt{{\cal R}(\zeta,v)}}\left[\widetilde{{\cal G}_1} \left(\frac{3 \Delta}{2 {\cal R}(\zeta,v)} \right) + 
\frac{{\cal R}^2(\zeta,0)}{{\cal R}^2(\zeta,v)}\widetilde{{\cal G}_2} \left(\frac{3 \Delta}{2 {\cal R}(\zeta,v)} \right)\right]\\
&&\hspace*{2cm}=\frac{2 \sqrt{2}}{\sqrt{3 \Delta}} \sinh^{-1}\left(\sqrt{\frac{3\Delta}{2}} \right) + \frac{\zeta - v}{\sqrt{6}},
\end{eqnarray}
which is valid up to terms of order ${\cal R}^{2}(0,\zeta)$, which is small when $\zeta \to \infty$ [see Eq. (\ref{expansion_R0})]. 
Hence from Eq. (\ref{explicit_expansion}), one expects that ${\cal R}(\zeta,v)$ admits the following expansion, for $\zeta \to \infty$, $v \to \infty$ with $v \sim \zeta$:
\begin{equation}\label{general_exp}
{\cal R}(\zeta,v) = H_0(\zeta,v) + {\cal R}^2(\zeta,0) H_1(\zeta,v) + {\cal O}\left({\cal R}^3(\zeta,0) \right) \;.
\end{equation}
As we will see, to obtain the asymptotic behavior of $p(\zeta)$ to lowest non-trivial order for large $\zeta$, we need to compute
${\cal R}(\zeta,v)$ up to order ${\cal R}^2(\zeta,0)$, i.e. we need to compute both $H_0(\zeta,v)$ and $H_1(\zeta,v)$ in Eq. (\ref{general_exp}).  

\vspace{0.5cm}

\noindent{\it Computation of $H_0(\zeta,v)$.} Neglecting the second term in the l.h.s. of Eq. (\ref{explicit_expansion}), 
one obtains the following equation for $H_0 \equiv H_0(\zeta,v)$:
\begin{equation}\label{eq_H0_1}
\frac{1}{\sqrt{H_0}} \widetilde{\cal G}_1\left(\frac{3\Delta}{2 H_0}\right) = \frac{2 \sqrt{2}}{\sqrt{3 \Delta}} \sinh^{-1}\left(\sqrt{\frac{3\Delta}{2}} \right) + \frac{\zeta - v}{\sqrt{6}} \;,
\end{equation}
which clearly shows that $H_0$ is a function of $\zeta-v$, $H_0(\zeta,v) = H_0(u=\zeta-v)$ where $H_0(u)$ is given by 
\begin{equation}\label{eq_H0_2}
H_0(u) = \frac{3\Delta}{2 \sinh^2{\left(\frac{\sqrt{\Delta}}{4} u + \sinh^{-1} \left( \sqrt{\frac{3\Delta}{2}}\right) \right)}} \;,
\end{equation}
which is obtained by injecting the explicit expression of $\widetilde{{\cal G}}_1(\alpha)$ in (\ref{def_G1}) into Eq. (\ref{eq_H0_1}). 

\vspace{0.5cm}

\noindent{\it Computation of $H_1(\zeta,v)$.}  
By inserting the expansion (\ref{general_exp}) into Eq. (\ref{explicit_expansion}) and expanding up to order ${\cal O}({\cal R}(\zeta,0)^2)$, 
one obtains that $H_1(\zeta,v)$ is also a function of $u = \zeta-v$ only, given by
\begin{eqnarray}
\label{def_H1}
\nonumber
H_1(u) &=& \frac{2}{3\Delta} Z(u) \frac{\widetilde{{\cal G}_2}(Z(u))}{\left[\frac{1}{2} 
\widetilde{{\cal G}_1}(Z(u)) + Z(u) \widetilde{{\cal G}_1}'(Z(u))\right]} \:,\;\\
&&{\rm with} \;\; Z(u) = \frac{3\Delta}{2 H_0(u)} \;,
\end{eqnarray}
where the functions $\widetilde{{\cal G}_1}(\alpha)$ and $\widetilde{{\cal G}_2}(\alpha)$ are given in 
Eq. (\ref{def_G1}) and Eq. (\ref{def_G2}) respectively and  
$\widetilde{{\cal G}_1}'(\alpha)= \frac{d}{d\alpha} \widetilde{{\cal G}_1}(\alpha)$.

\vspace{0.5cm}

\begin{figure}
\begin{center}
\includegraphics[width=1\linewidth]{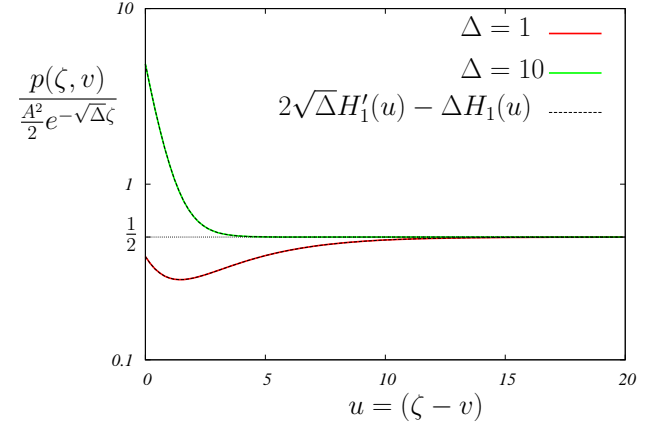}
\end{center}
\caption{Plot of the joint PDF $p(\zeta, v)$, obtained by numerically evaluating 
Eqs. (\ref{Rsv_subcritical}), (\ref{Rstar_subcritical}), (\ref{joint_PDF_sv}) and (\ref{margin_zeta}),
 as a function of $\zeta - v$, normalized by $\frac{A^2}{2} e^{-\sqrt{\Delta}\zeta}$ for a fixed $\zeta$ 
(which we choose to be a typical large value $\zeta = \exp(3)$).
The dashed lines denoting the function $2 \sqrt{\Delta} H_1'(\zeta-v) - \Delta H_1(\zeta-v)$ (derived in Eq. (\ref{joint_final})) 
for different $\Delta$ are indistinguishable from the theoretically obtained curves
as they match exactly. This function tends to $\frac{1}{2}$ as $\zeta-v \to \infty$, as predicted by Eq. (\ref{limit_H1}).}
\label{fig:largezeta_asymptotic}
\end{figure}

\noindent{\it Computation of the PDF of the span $p(\zeta)$ for large $\zeta$}. 
From Eq. (\ref{joint_PDF_sv_app}) together with the expansion in Eq. (\ref{general_exp}), we obtain the joint PDF $p(\zeta,v)$ as
\begin{eqnarray}
\label{joint_large_zeta}
\nonumber
&&\hspace{-0.5cm}p(\zeta,v) = \frac{1}{2}(\partial^2_v- \partial^2_\zeta) {\cal R}(\zeta,v) =\\
\nonumber
&&-\Big[ ({\cal R}'(\zeta,0))^2 + {\cal R}(\zeta,0){\cal R}''(\zeta,0)H_1(\zeta-v)\\
\nonumber
&&+ 2 {\cal R}(\zeta,0){\cal R}'(\zeta,0)H'_1(\zeta-v)  ) \Big] + {\cal O}({\cal R}^3(\zeta,0)) \;,\\
\end{eqnarray} 
where we have used $(\partial^2_v- \partial^2_\zeta) H_0(\zeta,v) = 0$ as $H_0(\zeta,v) = H_0(u=\zeta-v)$ depends only on $u = \zeta-v$. 
Using the large $\zeta$ expansion of ${\cal R}(\zeta,0)$ in Eq. (\ref{expansion_R0}), one finds, from (\ref{joint_large_zeta})
\begin{eqnarray}
\label{joint_final}
\nonumber
&&p(\zeta,v) = \frac{A^2}{2} e^{-\sqrt{\Delta}\zeta}\left(2 \sqrt{\Delta} H_1'(\zeta-v) - \Delta H_1(\zeta-v) \right)\\
&&\hspace{0.5\linewidth}+ {\cal O}\left(e^{-\frac{3\sqrt{\Delta}}{2}\zeta}\right)\;.
\end{eqnarray}
This behaviour of the joint PDF $p(\zeta,v)$ for large $\zeta$ is illustrated in Fig. \ref{fig:largezeta_asymptotic},
where we plot this distribution obtained by numerically evaluating Eqs. (\ref{Rsv_subcritical}), (\ref{Rstar_subcritical}) and (\ref{joint_PDF_sv}) 
for a fixed $\zeta$ (chosen to be a typical large value $\zeta = \exp(3)$).
We find a very good agreement between the exact PDF and the asymptotic behavior derived in Eq. (\ref{joint_final}) for all $\Delta$.
Finally, inserting this asymptotic behavior (\ref{joint_final}) of the joint PDF into Eq. (\ref{pzeta_app}) yields the PDF of the span which is given by
\begin{eqnarray}
\label{p_zeta_large_1}
\nonumber
&&p(\zeta) = A^2 e^{-\sqrt{\Delta}\zeta} \int_0^\zeta \left(2 \sqrt{\Delta} H_1'(u) - \Delta H_1(u) \right)\\
&&\hspace{0.5\linewidth}+ {\cal O}\left(e^{-\frac{3\sqrt{\Delta}}{2}\zeta}\right)\;.
\end{eqnarray}
Using Eq. (\ref{eq_H0_2}) together with Eq. (\ref{def_H1}), one can show, for instance using Mathematica that
\begin{equation}\label{limit_H1}
\lim_{u \to \infty}  \left(2 \sqrt{\Delta} H_1'(u) - \Delta H_1(u) \right) = \frac{1}{2} \;.
\end{equation}
This limiting behaviour (\ref{limit_H1}) is illustrated in Fig. \ref{fig:largezeta_asymptotic} for two different values $\Delta = 1$ and
$\Delta = 10$.
Finally, inserting this large $\zeta$ limiting behaviour (\ref{limit_H1}) into Eq. (\ref{p_zeta_large_1}), we arrive at
\begin{equation}
p(\zeta) = \frac{A^2}{2} \zeta  e^{-\sqrt{\Delta}~\zeta} \left(1 + {\cal O}(\zeta^{-1}) \right) \;,
\end{equation}
as given in Eq. (\ref{large_zeta_p_subcritical}) in the text.

\hspace{2cm}

\end{document}